\documentclass[english]{article}
\usepackage{lmodern}
\usepackage[T1]{fontenc}
\usepackage[utf8]{inputenc}
\usepackage[a4paper]{geometry}
\usepackage{float}
\usepackage{textcomp}
\usepackage{mathtools}
\usepackage{amsmath}
\usepackage{graphicx}
\usepackage{microtype}

\makeatletter

\providecommand{\tabularnewline}{\\}
\floatstyle{ruled}
\newfloat{algorithm}{tbp}{loa}
\providecommand{\algorithmname}{Algorithm}
\floatname{algorithm}{\protect\algorithmname}

\newcommand{\lyxaddress}[1]{
	\par {\raggedright #1
	\vspace{1.4em}
	\noindent\par}
}

\usepackage[unicode=true, bookmarks=true,bookmarksnumbered=false,bookmarksopen=false, breaklinks=false,pdfborder={0 0 0},pdfborderstyle={},backref=false,colorlinks=false]{hyperref}
\hypersetup{pdftitle={Auto-Calibration of Cone Beam Geometries from Arbitrary Rotating Markers using a Vector Geometry Formulation of Projection Matrices},pdfauthor={Jonas Graetz}}
\usepackage{algpseudocode}
\newcommand{\algvspace}{\vspace{\parsep}}
\usepackage{setspace}
\usepackage{multicol}
\usepackage[font={small,sf},labelfont=bf]{caption}
\usepackage{microtype}
\usepackage{fancyhdr}
\usepackage{scalerel}
\usepackage{tikz}
\usetikzlibrary{svg.path}

\tikzset{
  orcidlogo/.pic={
    \fill[gray] svg{M256,128c0,70.7-57.3,128-128,128C57.3,256,0,198.7,0,128C0,57.3,57.3,0,128,0C198.7,0,256,57.3,256,128z};
    \fill[white] svg{M86.3,186.2H70.9V79.1h15.4v48.4V186.2z}
                 svg{M108.9,79.1h41.6c39.6,0,57,28.3,57,53.6c0,27.5-21.5,53.6-56.8,53.6h-41.8V79.1z M124.3,172.4h24.5c34.9,0,42.9-26.5,42.9-39.7c0-21.5-13.7-39.7-43.7-39.7h-23.7V172.4z}
                 svg{M88.7,56.8c0,5.5-4.5,10.1-10.1,10.1c-5.6,0-10.1-4.6-10.1-10.1c0-5.6,4.5-10.1,10.1-10.1C84.2,46.7,88.7,51.3,88.7,56.8z};
  }
}

\newcommand\orcidicon[1]{\href{https://orcid.org/#1}{\mbox{\scalerel*{
\begin{tikzpicture}[yscale=-1,transform shape]
\pic{orcidlogo};
\end{tikzpicture}
}{|}}}}

\makeatother

\usepackage{babel}
\begin{document}
\title{Auto-Calibration of Cone Beam Geometries \\
from Arbitrary Rotating Markers using a\\
Vector Geometry Formulation of Projection Matrices}
\author{Jonas Graetz (Dittmann) \textsuperscript{\orcidicon{0000-0002-4403-3686}}}
\maketitle

\lyxaddress{\begin{center}
\textit{Lehrstuhl für Röntgenmikroskopie, Physikalisches Institut,
Universität Würzburg }\\
\textit{Fraunhofer IIS, Division EZRT, Department MRB}\\
\textit{Josef-Martin-Weg 63, 97074 Würzburg, Germany}
\par\end{center}}
\begin{abstract}
A method for the determination of the projection geometry of highly
magnifying cone beam micro computed tomography systems based on few
rotating fiducial markers of unknown position within the field of
view is derived. By employing the projection matrix formalism commonly
used in computer graphics, a very clear presentation of the resulting
self consistent calibration problem can be given relating the sought-for
matrix to observable parameters of the markers' projections. Both
an easy to implement solution procedure for both the unknown projection
matrix and the marker assembly as well as the mapping from projection
matrices to real space positions and orientations of source and detector
relative to the rotational axis are provided. 

The separate treatment of the calibration problem in terms of projection
matrices on the one hand and the independent transformation to a more
intuitive geometry representation on the other hand proves to be very
helpful with respect to the discussion of the ambiguities occurring
in reference-free calibration. In particular, a link between methods
based on knowledge on the sample and those based on knowledge solely
on the detector geometry can be drawn. This further provides another
intuitive view on the often reported difficulty in the estimation
of the detector tilt towards the rotational axis.

A simulation study considering $10^{6}$ randomly generated cone beam
imaging configurations and fiducial marker distributions within a
range of typical scenarios is performed in order to assess the stability
of the proposed technique. 
\end{abstract}
\thispagestyle{fancy}

\section{Introduction}

The determination of the actual – in contrast to the intended – projection
geometry within an X-ray cone beam computed tomography system is a
common problem to be solved prior to tomographic volume reconstruction.
In particular for high resolution micro computed tomography systems,
due to the very small source–object distances required for large magnification
factors, external determination of the exact geometry is usually difficult
given its high sensitivity to small variations in the sample position
with respect to the source. It is for this reason generally desirable
to determine the actual projection geometry using the system's immanent
imaging capabilities, and most commonly, sparse assemblies of opaque
markers that can be individually segmented within projections thereof
are used for this purpose.

\subsection{Motivation and contribution of the present work}

The objective of the present work is to present a simple solution
to the cone beam CT auto-calibration problem based on rotating fiducial
markers of unknown position on the one hand and to provide intuitive
insights into the problem, its solution and its limitations on the
other hand. In contrast to the previous work on CBCT auto-calibration,
the calibration problem will be formulated in terms of projection
matrix elements with the problem of conversion between projection
matrices and real space geometry definition being treated separately.
It will turn out that this leads to a very clear view on the calibration
problem avoiding the cumbersome trigonometric relations between classic
geometry parametrizations and the perspective projections of points
in 3D space. In particular, the use of generic least squares optimization
is largely avoided. The consideration of constraints (i.e., prior
knowledge) on either the calibration structure or the projection system
takes the form of homography transformations and is separated as an
individually solvable problem, rendering the method more general.
This allows straight forward extensions to different kinds of prior
knowledge, which becomes a matter of constraining the remaining degrees
of freedom in the homography matrix, without touching the initial
procedure solving the calibration equations.

The present article is structured as follows: First, an overview over
related literature on marker based geometry calibration will be given,
providing context and motivation for the present work. A very brief
introduction to homogeneous coordinates and the projection matrix
formalism will be given. A small set of linear equations relating
both the unknown projection matrix elements and the parameters of
the circular marker trajectories to observable parameters emerges
almost directly from the projection matrix based forward model of
the projection of rotating points. The following section introduces
transformations between projection matrices and real space vector
geometry, which provide important insights with respect to constraining
the ambiguities arising within auto-calibration. Explicit algorithms
implementing the derived strategies are given in the appendix. Finally,
a simulation study will be presented, assessing the stability of the
procedure on noisy projections of randomly generated sets of projection
geometries and marker placements.

\subsection{Literature review}

\subsubsection{Classification of different approaches}

Two general classes of approaches can be identified: those based on
precisely defined or known marker assemblies \cite{RougeeTrousset1993,Rizo1994,DeMenthonDavis1995,NavabGraumann1996,StrobelWiesent2003,Cho2005,StrubelNoo2005,KYangBoone2006,HoppeNooHornegger2007,MaoXing2008,JohnstonBadea2008,RobertMainprize2009,MennessierClackdoyleNoo2009,XLiBLiu2010,FordWilliamson2011,MXuDLi2014,ZechnerSteiniger2016},
which allow calibration on a per-view basis, and those working with
fiducials of unknown placement yet requiring precise circular motion
of these markers throughout a tomographic scan \cite{Gullberg1990,JLiColeman1993,Noo2000,Smekal2004,KYangBoone2006,WangTsui2007,DWuYXiao2011,GrossSchwanecke2012,SawallKachelriess2012,XuTsui2013,GLi2019}
(with some of these assuming the distances between markers known \cite{Noo2000,KYangBoone2006,WangTsui2007,DWuYXiao2011,XuTsui2013}).
While the former methods are typically used for macroscopic systems
with fields of view in the range of 10 cm and larger, the latter are
required for microscopic systems for which the manufacturing of well-defined
calibration phantoms is hard to impossible. \cite{Beque2003} assumes
both known markers and perfect rotation, and \cite{DefriseNuyts2008}
provides a method to account for deviations from expected precise
motions, enabling per-view calibration also for methods originally
assuming stable circular motions.

An additional distinction can be made from the technical point of
view of system parametrization: methods aiming to determine the projective
mapping from 3D to 2D space in terms of a projection matrix that is
consistent with the available observations irrespective of the question
how the particular mapping arises physically \cite{RougeeTrousset1993,NavabGraumann1996,StrobelWiesent2003,StrubelNoo2005,HoppeNooHornegger2007,XLiBLiu2010},
and methods aiming to relate projections or properties thereof to
real space geometry parameters (relative distances and orientation
angles of source and detector) \cite{Gullberg1990,JLiColeman1993,RougeeTrousset1993,Rizo1994,Noo2000,Beque2003,Smekal2004,Cho2005,KYangBoone2006,MaoXing2008,JohnstonBadea2008,RobertMainprize2009,MennessierClackdoyleNoo2009,FordWilliamson2011,DWuYXiao2011,GrossSchwanecke2012,SawallKachelriess2012,XuTsui2013,MXuDLi2014,GLi2019}.
Differences also exist in the evaluation of the projection data used
for calibration: methods directly working on extracted projection
samples (2D points) without further data reduction or interpretation
\cite{Gullberg1990,JLiColeman1993,RougeeTrousset1993,DeMenthonDavis1995,NavabGraumann1996,Beque2003,StrobelWiesent2003,HoppeNooHornegger2007,MaoXing2008,SawallKachelriess2012,ZechnerSteiniger2016,GLi2019},
as well as methods reducing the observed projections by means of matching
them to an expected model (such as e.g.\ elliptic trajectories)
or otherwise exploiting specific geometric features of the utilized
calibration structure \cite{Rizo1994,Noo2000,Smekal2004,Cho2005,StrubelNoo2005,KYangBoone2006,JohnstonBadea2008,RobertMainprize2009,MennessierClackdoyleNoo2009,FordWilliamson2011,DWuYXiao2011,GrossSchwanecke2012,XuTsui2013,MXuDLi2014}.
Calibration methods may further be characterized based on their core
calibration approaches: \cite{RougeeTrousset1993,NavabGraumann1996,StrobelWiesent2003,StrubelNoo2005,HoppeNooHornegger2007,XLiBLiu2010}
reduce the calibration problem to the solution of a linear system
of equations in a least squares sense e.g.\ by means of singular
value decomposition (requiring the imaged object to be known). \cite{Noo2000,Smekal2004,Cho2005,KYangBoone2006,MennessierClackdoyleNoo2009,XuTsui2013,MXuDLi2014}
derive direct relations between parameters of the observed marker
patterns and the underlying projection geometry. \cite{Gullberg1990,JLiColeman1993,RougeeTrousset1993,Beque2003,WangTsui2007,ZechnerSteiniger2016}
use local optimization techniques requiring sufficiently good initial
estimates and \cite{MaoXing2008,GrossSchwanecke2012,SawallKachelriess2012,GLi2019}
rely on global optimization techniques for the solution of the inverse
problem (also requiring initial estimates). Combinations are used
e.g.\ in \cite{Rizo1994,DeMenthonDavis1995,JohnstonBadea2008,RobertMainprize2009},
and a comparison of a matrix inversion and an optimization based approach
is given by \cite{RougeeTrousset1993}. Finally, the cited methods
differ in the amount of degrees of freedom that are addressed. While
methods based on fully known reference objects are generally able
to determine all system parameters with reasonable precision, the
situation is more complex when no or only few assumptions can be made
on the calibration structure. 

\subsubsection{Review of previous auto-calibration techniques}

The early methods addressing the problem of calibration from unknown
markers with straight forward least squares minimization approaches
considered only the focal distance and the most relevant translations
of the detector and the rotational axis \cite{Gullberg1990,JLiColeman1993}
due to the instability of the full optimization problem. Noo et al.\
\cite{Noo2000} showed that generic optimization can be largely avoided
by systematic analysis of the problem. By constraining the detector
columns parallel to the rotational axis, they were able to derive
relations between the remaining geometric parameters and the ellipse
parameters of the observable projections of two opaque markers moving
along circular trajectories around the rotational axis. Potential
detector in-plane rotations are considered in an independent preprocessing
step. They assumed the distance between the markers known in order
to determine also the absolute scale. The method was further simplified
by Yang et al\@.\ \cite{KYangBoone2006} for the case of also negligible
detector slant about the rotational axis. Johnston et al.\ \cite{JohnstonBadea2008}
use the latter approximate approach for the initialization of a generic
local optimization procedure that is then able to recover all parameters
using a phantom of ten collinear bearing balls with defined distances.
A further adaption of the Noo method was presented by \cite{DWuYXiao2011}.
Bequé et al.\ \cite{Beque2003,Beque2005} and Wang and Tsui \cite{WangTsui2007}
fully revert to local optimization again yet analyze the uniqueness
of general solutions based on one, two and three rotating markers
and conclude that one known distance between two markers is sufficient
in the presence of detector slant about the rotational axis, and two
known distances (which they relate to distances between three markers)
are required for a generally unique solution for all system parameters,
which was also conjectured previously by Noo et al.\ \cite{Noo2000}
and is also stated by Xu and Tsui \cite{XuTsui2013}. In the related
methods addressing all system parameters based on two known parallel
rings (of equal radius) of bearing balls by Cho et al.\ \cite{Cho2005}
and Robert et al.\ \cite{RobertMainprize2009}, this condition is
fulfilled by knowledge of both the vertical distance of those circles
and their (common) radius. Further Xu and Tsui \cite{XuTsui2013}
propose another calibration procedure based on relations between elliptical
projections of rotating markers, known marker distances and the system
geometry, yet in contrast to \cite{Noo2000,KYangBoone2006,RobertMainprize2009}
identify geometrically meaningful intersection points on the ellipses
in the style of the approaches by Cho et al.\ \cite{Cho2005} and
Strubel et al.\ \cite{StrubelNoo2005} (who worked with fully known
structures) in order to then obtain simpler relations between those
points in the projection image and the geometry parameters. A general
methodology for the assessment of uniqueness and stability of calibration
problems by means of analyzing the propagation of random errors through
the respective forward model has been discussed by Ma et al.\ \cite{TMa2009}.

Smekal et al.\ \cite{Smekal2004} were, to the author's knowledge,
the first to explicitly address the case of complete auto-calibration
(up to an unknown object scale, yet including detector tilt) of cone
beam tomography systems based on multiple rotating markers at unknown
positions \emph{and distances}. The latter were either required or
assumed known in the methods discussed so far. They choose a different
parametrization for the analysis of the projected circles more directly
related to the forward model than the otherwise often used ellipse
equation and are able to find relations of those observable parameters
to all geometric parameters. In contrast to the previous literature,
Smekal et al\@. conclude that one marker is, in principle, sufficient
for all parameters but tilt (and absolute scale). In accordance with
previous literature they find that at least two projected trajectories
(yet without knowledge of the marker's distance) are required to infer
tilt provided that the detector is also slanted about the rotational
axis. They do not further investigate the case of zero slant. In their
experiments,  they use between eight and twelve markers and average
the obtained geometry parameters.

The more recent publications by Gross et al.\ \cite{GrossSchwanecke2012},
Sawall et al.\ \cite{SawallKachelriess2012} and Li et al.\ \cite{GLi2019}
also consider the problem of complete auto-calibration (up to an unknown
scale) without reverting to known sample properties, although all,
in contrast to Smekal, require global optimization techniques. 

Gross et al.\ \cite{GrossSchwanecke2012} formulate the problem in
terms of a homography transform parametrized by the sought-for system
properties relating the observable ellipses to a canonical representation
of circular trajectories. They were thereby able to eliminate the
trajectory parameters from the optimization problem, reducing it to
6 degrees of freedom describing the projection geometry (in comparison,
Robert et al.\ \cite{RobertMainprize2009} previously reported a
reduction of the optimization problem relating ellipse parameters
and geometry parameters to 3 dimensions for the case of known trajectory
parameters). As has been the case also with previous techniques based
on ellipse analysis \cite{Noo2000,Cho2005,StrubelNoo2005,KYangBoone2006,RobertMainprize2009,DWuYXiao2011,XuTsui2013},
the evaluated ellipses are expected to be non-degenerate in order
to successfully reconstruct the projection geometry. In their experiments,
they use between three and twelve markers and generally recommend
the use of more than four non-degenerate ellipses for their method.

Sawall et al.\ \cite{SawallKachelriess2012} apply a genetic optimization
algorithm to a straight-forward objective function penalizing the
least squares errors between forward model and observations. Rather
than using multiple rotating markers within one tomographic scan,
they use a single marker scanned at multiple different geometric configurations
of the employed tomography setup. As discussed earlier \cite{Beque2003,WangTsui2007},
this provides enough information to simultaneously determine each
scan geometry. Sawall et al.\ were the only ones to actually work
with the absolute minimum of one marker, although simultaneous calibration
of multiple systems or multiple configurations of the same system
has been addressed before \cite{Beque2003,WangTsui2007,JohnstonBadea2008}.

Li et al.\ \cite{GLi2019} likewise use global optimization to relate
the forward model to observed marker projections, yet specifically
design the cost function to exploit known consistency constraints
of two-view geometries. Namely, the lines between focal spot, marker
and the marker's projection on the detector for two views of the same
marker must obviously intersect, and do so at the location of the
marker. The complete elimination of degrees of freedom to be optimized,
as previously shown by Gross et al.\ \cite{GrossSchwanecke2012},
could thereby not be achieved.

In the following, an analytically motivated approach to reference-free
calibration will be presented that, in contrast to the ones by Gross
et al.\ \cite{GrossSchwanecke2012}, Sawall et al.\ \cite{SawallKachelriess2012}
and Li et al.\ \cite{GLi2019}, does not require the use of generic
non-convex optimization techniques, and thus in particular does not
require initial estimates for any of the parameters. It is in this
respect most similar to the method described by Smekal et al.\ \cite{Smekal2004}.
Other than in \cite{Smekal2004} and other previous approaches, the
problem will be parametrized by projection matrix elements, which
has previously only been done in the context of known calibration
structures (cf.\ \cite{RougeeTrousset1993,NavabGraumann1996,StrobelWiesent2003,StrubelNoo2005,HoppeNooHornegger2007,XLiBLiu2010}).
This will on the one hand allow a very simple representation of the
core calibration problem and on the other hand expose the link between
methods solely based on constraints on the detector geometry (\cite{Smekal2004,GrossSchwanecke2012,SawallKachelriess2012,GLi2019})
and those based on (partial) prior knowledge on the sample (e.g.\
\cite{Beque2003,Beque2005,StrubelNoo2005,Cho2005,WangTsui2007,JohnstonBadea2008,RobertMainprize2009,XuTsui2013}).

\section{Methods}

\subsection{Formalization of the calibration task\label{subsec:pmat-model}}

\begin{figure}
\begin{centering}
\includegraphics[viewport=10bp 10bp 430bp 346bp,width=0.48\textwidth]{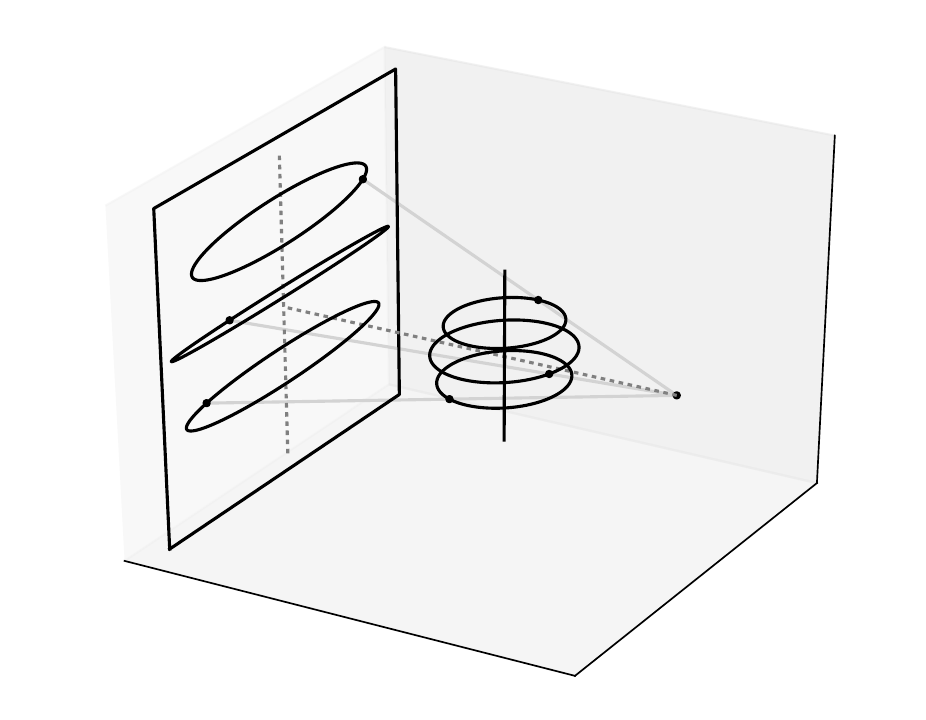}%
\begin{minipage}[b][1\totalheight][c]{0.385\textwidth}%
\begin{center}
\includegraphics[width=1\textwidth]{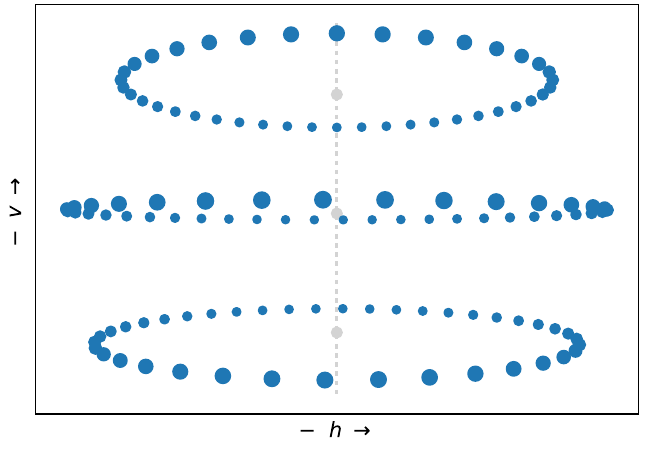}\vspace{2em}
\par\end{center}%
\end{minipage}
\par\end{centering}
\begin{centering}
~~~~~~~\includegraphics[width=0.385\textwidth]{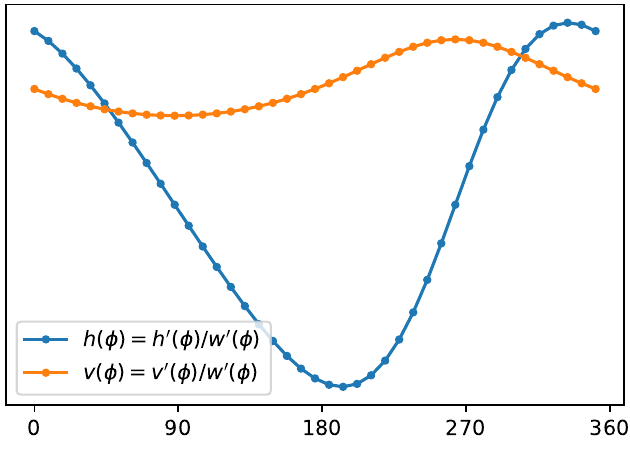}~~~\includegraphics[width=0.385\textwidth]{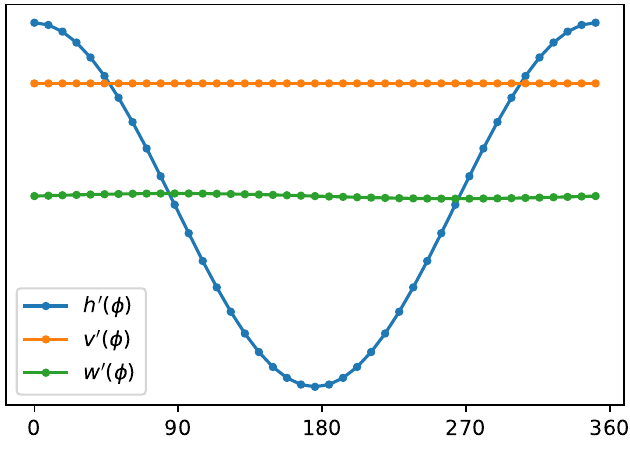}
\par\end{centering}
\caption{\label{fig:projection-demo}\emph{Upper left}: Fiducial markers moving
along circular trajectories about the rotational axis (center) of
a cone beam computed tomography setup, projected by a point source
(right) onto a planar detector (left). $\quad$\emph{Upper right}:
Superposition of respective projection images on the detector. $\quad$\emph{Lower
left}: Horizontal and vertical components of a projected trajectory
in dependence of the projection angle $\phi$ (cf.\ Eqs.\ \ref{eq:projection_perspective_h},
\ref{eq:projection_perspective_v}). $\quad$\emph{Lower right}: Decomposition
of the projection components into independent sinusoids describing
the orthographic horizontal and vertical projections $h'(\phi)$ and
$v'(\phi)$ as well as the perspective scaling component $w'(\phi)$
(cf.\ Eq.~\ref{eq:projection_homogeneus_form}).}
\end{figure}
\begin{figure}

\centering{}\includegraphics[viewport=30bp 30bp 431bp 326bp,width=0.46\textwidth]{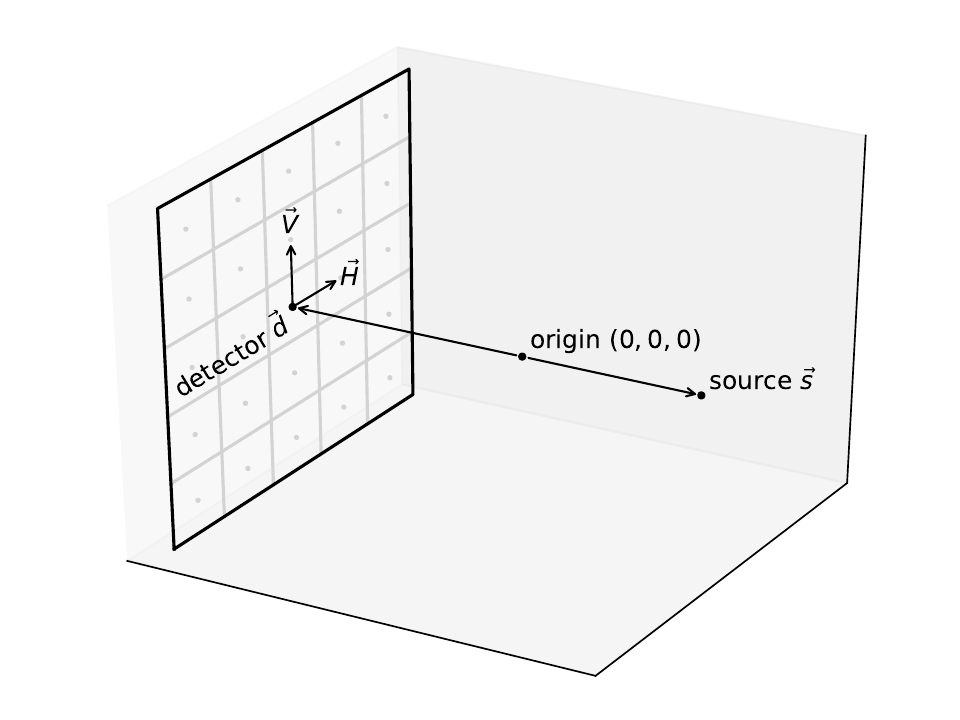}\caption{\label{fig:geometry-def}Depiction of the real space geometry description
used here. The world coordinate origin is centered in the field of
view and the positions of source ($\vec{s}$) and detector coordinate
origin $(\vec{d}$) are defined relative to that. The row and column
vectors $\vec{H}$ and $\vec{V}$ characterize the detector orientation
and pixel pitches. This vectorial description does not restrict $\vec{H}$
and $\vec{V}$ to be orthogonal or of equal length, although this
is a common property of actual imaging systems and will here be used
to constrain ambiguities of the auto calibration problem. The 3D world
coordinates of a 2D point $(h,v)$ in the planar coordinate system
of the detector are given by $\vec{d}+h\vec{H}+v\vec{V}$. The relation
to projection matrices is described in Section~\ref{subsec:pmat-to-realspace}.}
\end{figure}
Perspective projections onto planar detectors can generally be expressed
in terms of homogeneous coordinates and projection matrices as commonly
used in computer graphics. Homogeneous coordinates hold an additional
scaling component (here:\ $w$) defining an equivalence class such
that all vectors $(w\,x,\,w\,y,\,w\,z,\,w)$ with $w\neq0$ describe
the same point $(x,y,z)$ \cite{HartleyZisserman2004}. Corresponding
euclidean coordinates are obtained by dividing a given homogeneous
vector by its scaling component and dropping the latter. While the
motivation for this representation will become more clear in Section~\ref{subsec:pmat-to-realspace},
it is for now only relevant to know that the 2D cone beam projection
$(h(\phi),v(\phi))$ of a circular trajectory $(r\cos(\phi-\phi_{0}),r\sin(\phi-\phi_{0}),z)$
about the $z$-axis can generally be represented as:
\begin{align}
\left[\begin{array}{c}
h'(\phi)\\
v'(\phi)\\
w'(\phi)
\end{array}\right] & =\left[\begin{array}{cccc}
P_{11} & P_{12} & P_{13} & P_{14}\\
P_{21} & P_{22} & P_{23} & P_{24}\\
P_{31} & P_{32} & P_{33} & P_{34}
\end{array}\right]\left[\begin{array}{c}
r'\cos(\phi-\phi_{0})\\
r'\sin(\phi-\phi_{0})\\
z'\\
w
\end{array}\right]\\
h(\phi) & =h'(\phi)/w'(\phi)\\
v(\phi) & =v'(\phi)/w'(\phi)
\end{align}
with 
\begin{align*}
r & =r'/w\\
z & =z'/w\,.
\end{align*}
$w$ and $w'$ are the scaling components of the vector on the right-hand
side and its projection on the left-hand side respectively. Although
the component $w$ on the right-hand side would, if projections of
known points were to be calculated, commonly be defined to equal $1$,
it will be beneficial for the purposes of the following derivations
to actually leave it as an open parameter. $P_{mn}$ represent the
components of a $3\times4$ projection matrix $\boldsymbol{P}$ encoding
the projection geometry, i.e.\ the placement of source and detector
in 3D space as well as the detector pixel pitches. The detailed relation
between geometry and projection matrix will be discussed later in
Section~\ref{subsec:pmat-to-realspace}.

Evaluating the above matrix-vector product and applying trigonometric
identities yields

\begin{align}
\left[\begin{array}{c}
h'(\phi)\\
v'(\phi)\\
w'(\phi)
\end{array}\right] & =\left[\begin{array}{c}
r'(P_{11}\cos(\phi-\phi_{0})+P_{12}\sin(\phi-\phi_{0}))+P_{13}z'+P_{14}w\\
r'(P_{21}\cos(\phi-\phi_{0})+P_{22}\sin(\phi-\phi_{0}))+P_{23}z'+P_{24}w\\
r'(P_{31}\cos(\phi-\phi_{0})+P_{32}\sin(\phi-\phi_{0}))+P_{33}z'+P_{34}w
\end{array}\right]\label{eq:projection_homogeneus_form}\\
 & =\left[\begin{array}{c}
r'P_{1\mathrm{a}}\sin(\phi-\phi_{0}-\phi_{\mathrm{h}})+P_{13}z'+P_{14}w\\
r'P_{2\mathrm{a}}\sin(\phi-\phi_{0}-\phi_{\mathrm{v}})+P_{23}z'+P_{24}w\\
r'P_{3\mathrm{a}}\sin(\phi-\phi_{0}-\phi_{\mathrm{w}})+P_{33}z'+P_{34}w
\end{array}\right]
\end{align}
with 
\begin{align*}
P_{m\mathrm{a}} & =\sqrt{P_{m1}^{2}+P_{m2}^{2}}\\
\phi_{\mathrm{h}} & =\mathrm{arctan}2(-P_{11},P_{12})\\
\phi_{\mathrm{v}} & =\mathrm{arctan}2(-P_{21},P_{22})\\
\phi_{\mathrm{w}} & =\mathrm{arctan}2(-P_{31},P_{32})\\
 & \Updownarrow\\
P_{m1} & =-P_{1\mathrm{a}}\sin(\phi_{m})\\
P_{m2} & =\phantom{-{}}P_{1\mathrm{a}}\cos(\phi_{m})\:,
\end{align*}
where $m$ equally enumerates both the rows $(1$, $2$ and $3$)
of $\boldsymbol{P}$ as well the associated subscript labels $\mathrm{h}$,
$\mathrm{v}$ and $\mathrm{w}$, which have been chosen to clearly
indicate the relation of the rows or their respective parameters to
the three components of the left-hand side homogeneous vector.

The $h$ and $v$ projection coordinates on the detection plane are
thus finally given by:
\begin{align}
h(\phi) & =\frac{r'P_{1\mathrm{a}}\sin(\phi-\phi_{0}-\phi_{\mathrm{h}})+P_{13}z'+P_{14}w}{r'P_{3\mathrm{a}}\sin(\phi-\phi_{0}-\phi_{\mathrm{w}})+P_{33}z'+P_{34}w}\label{eq:projection_perspective_h}\\
v(\phi) & =\frac{r'P_{2\mathrm{a}}\sin(\phi-\phi_{0}-\phi_{\mathrm{v}})+P_{23}z'+P_{24}w}{r'P_{3\mathrm{a}}\sin(\phi-\phi_{0}-\phi_{\mathrm{w}})+P_{33}z'+P_{34}w}\quad.\label{eq:projection_perspective_v}
\end{align}
The numerators in these expressions correspond to the orthographic
(``parallel beam'') projection, while the common denominator describes
the distance dependent perspective scaling. See Figure~\ref{fig:projection-demo}
for an example.

As in the context of self consistent calibration neither $P_{mn}$
nor $r'$, $z'$, $\phi_{0}$ and $w$ are known, the above equations
may as well be expressed in terms of the independent sinusoid parameters
$a_{i\mathrm{h}}$, $a_{i\mathrm{v}}$, $a_{i\mathrm{w}}$, $\phi_{i0\mathrm{h}}$,
$\phi_{i0\mathrm{v}}$, $\phi_{i0\mathrm{w}}$, $o_{i\mathrm{h}}$
and $o_{i\mathrm{v}}$ representing amplitudes, phases and offsets
respectively and including an additional index $i$ for the enumeration
of multiple projected trajectories:
\begin{align}
h_{i}(\phi) & =\frac{a_{i\mathrm{h}}\sin(\phi-\phi_{i0\mathrm{h}})+o_{i\mathrm{h}}}{a_{i\mathrm{w}}\sin(\phi-\phi_{i0\mathrm{w}})+1}\label{eq:sinusoid-fw-model_h}\\
v_{i}(\phi) & =\frac{a_{i\mathrm{v}}\sin(\phi-\phi_{i0\mathrm{v}})+o_{i\mathrm{v}}}{a_{i\mathrm{w}}\sin(\phi-\phi_{i0\mathrm{w}})+1}\quad.\label{eq:sinusoid-fw-model_v}
\end{align}
The determination of these 8 sinusoid parameters for each projected
trajectory $i$ is detailed in Section~\ref{subsec:parameter-extraction}
(Eqs.\ \ref{eq:sinusoid-params-start}–\ref{eq:sinusoid-params-mean-c_w}
and Algorithm~\ref{alg:pmat-recon-sinusoids}). In the following,
they can be considered known.

The calibration problem, i.e.\ the simultaneous reconstruction of
both the unknown projection matrix $\boldsymbol{P}$ and the unknown
fiducial marker orbits from given projections $(h_{i}(\phi),v_{i}(\phi)$),
can now be identified as the solution of the following linear system
of equations relating the observable sinusoid parameters of the projected
trajectories to the unknown projection matrix and orbit parameters
for all imaged trajectories $i$:
\begin{align}
r_{i}'P_{1\mathrm{a}} & =a_{i\mathrm{h}}\label{eq:calib-a_h}\\
r_{i}'P_{2\mathrm{a}} & =a_{i\mathrm{v}}\label{eq:calib-a_v}\\
r_{i}'P_{3\mathrm{a}} & =a_{i\mathrm{w}}\label{eq:calib-a_w}\\
\nonumber \\
P_{13}z_{i}'+P_{14}w_{i} & =o_{i\mathrm{h}}\label{eq:calib-o_h}\\
P_{23}z_{i}'+P_{24}w_{i} & =o_{i\mathrm{v}}\label{eq:calib-o_v}\\
P_{33}z_{i}'+P_{34}w_{i} & =o_{i\mathrm{w}}=1\label{eq:calib-o_w}\\
\nonumber \\
\phi_{\mathrm{h}} & =0\\
\phi_{\mathrm{v}} & =\phi_{i0\mathrm{v}}-\phi_{i0\mathrm{h}}\\
\phi_{\mathrm{w}} & =\phi_{i0\mathrm{w}}-\phi_{i0\mathrm{h}}\label{eq:calib-phi_w}
\end{align}
where $\phi_{\mathrm{h}}$ is defined to be $0$, exploiting the freedom
of choice of the projection angle for a rotationally symmetric imaging
configuration (or equivalently the freedom of choice of the initial
phase of a periodic trajectory). Further, the fixed parameter $o_{i\mathrm{w}}=1$
has been introduced in order to maintain a uniform representation
of the equations\@. For the same reason, the subscripts ``h'',
``v'' and ``w'' will in the following as well be represented by
the projection matrix' row index $m$, i.e.\ $\mathrm{h}\,\widehat{=}\,1,\mathrm{v}\,\widehat{=}\,2,\mathrm{w}\,\widehat{=}\,3$.
A practical method for the solution of the above calibration equations
is described in Section~\ref{subsec:iterative-pmat-recon}.

\subsection{Projective ambiguties\label{subsec:ambiguities}}

\begin{figure}

\centering{}\includegraphics[viewport=20bp 0bp 455bp 336bp,width=0.4\textwidth]{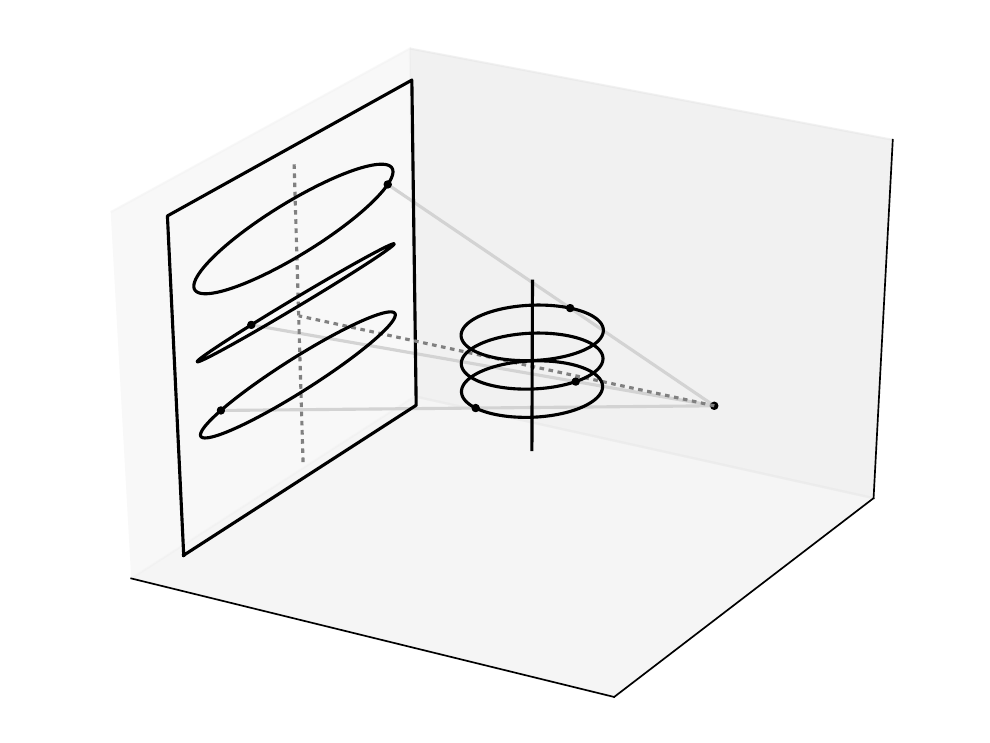}~~\includegraphics[viewport=20bp 0bp 455bp 336bp,width=0.4\textwidth]{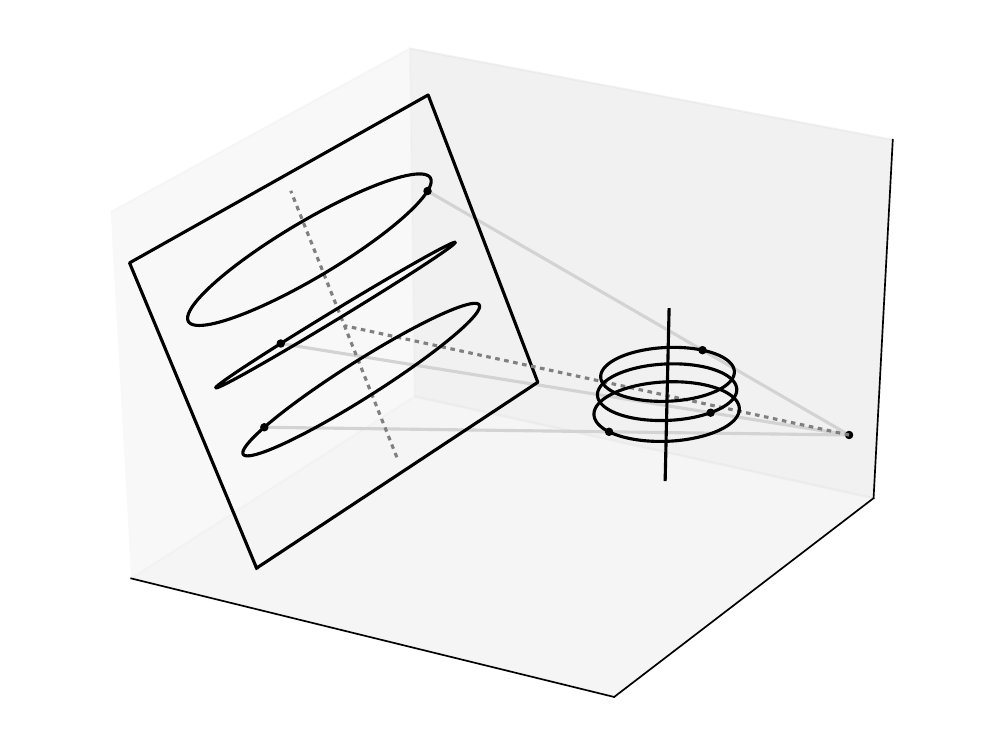}\\
\includegraphics[viewport=20bp 20bp 455bp 336bp,width=0.4\textwidth]{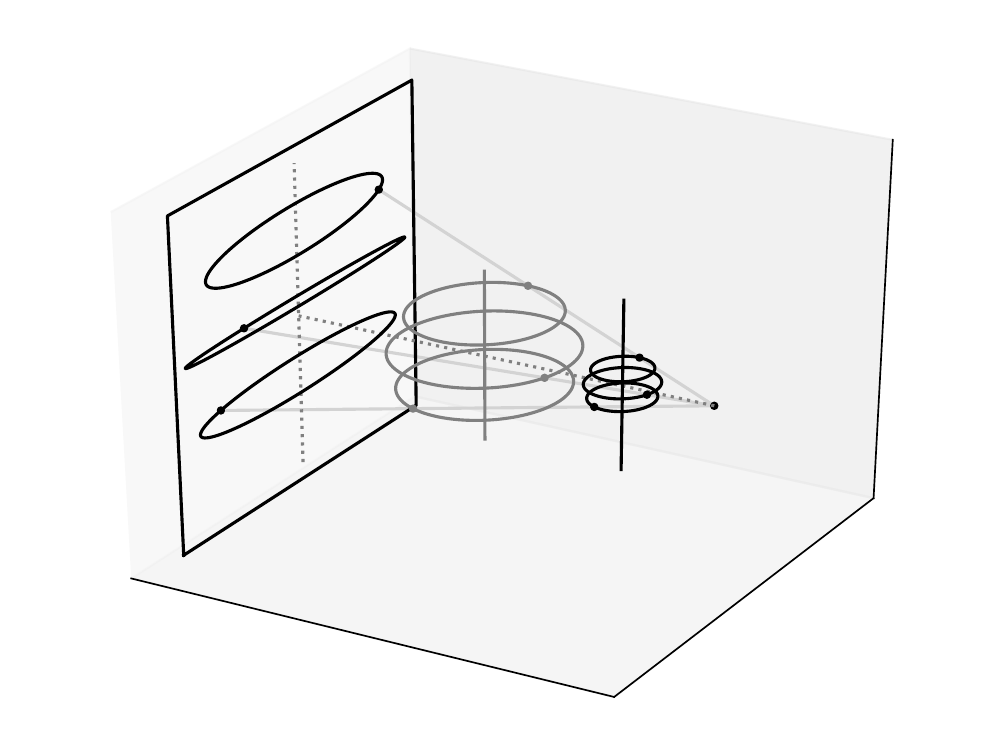}~~\includegraphics[viewport=20bp 20bp 455bp 336bp,width=0.4\textwidth]{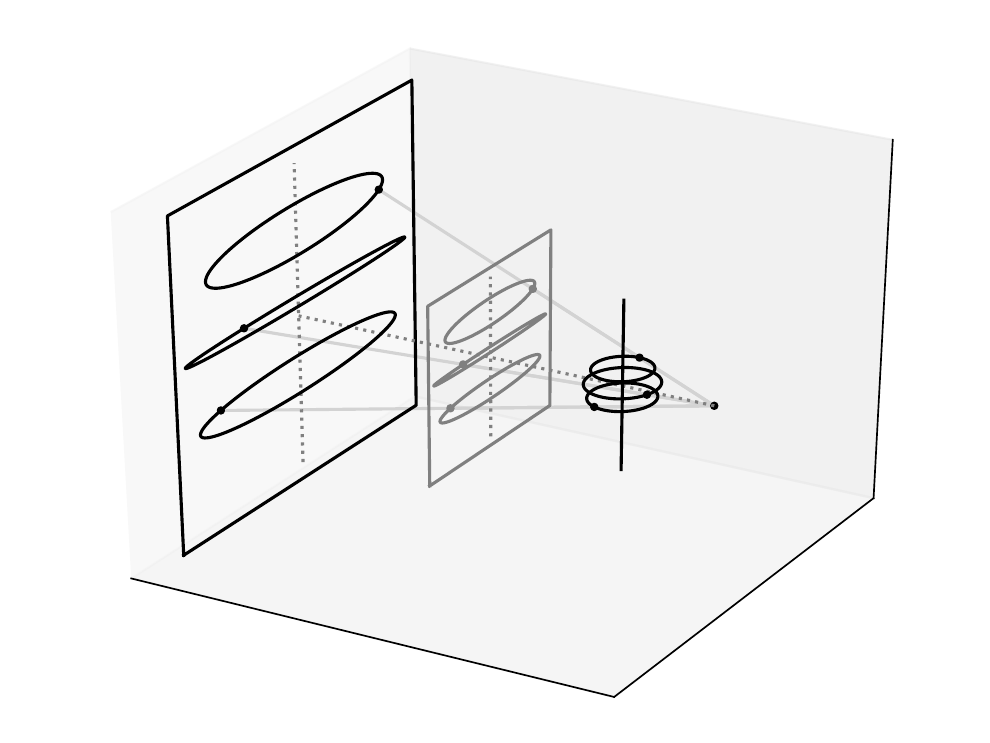}\caption{\label{fig:projective-ambiguity}Examples of projective ambiguity:
the same projections $(h(\phi),v(\phi))$ may originate from different
combinations of imaging configuration and marker locations. The top
row illustrates the effect of $\boldsymbol{D}(\delta=1,\gamma\protect\neq0)$,
while the bottom row shows examples of $\boldsymbol{S}$ with the
effects of the parameters $s_{\mathrm{o}}$ (bottom left) and $s_{\mathrm{d}}$
(bottom right) respectively (cf. Eqs.~\ref{eq:general-projective-ambiguities}–\ref{eq:delta-gamma-homography}).
The latter case corresponds to the general scaling degree of freedom
of homogeneous projection matrices.}
\end{figure}
The given equations further reveal that the first two and last two
columns of $\boldsymbol{P}$ are independent, i.e.\ there are no
equations interrelating these parts of the projection matrix (Eqs.~\ref{eq:calib-a_h}–\ref{eq:calib-a_w}
are independent from Eqs.~\ref{eq:calib-o_h}–\ref{eq:calib-o_w}).
Similarly, the relation between $z_{i}'$ and $w_{i}$ and consequently
the third and fourth column of $\boldsymbol{P}$ is not unique (cf.
Eqs.~\ref{eq:calib-o_h}–\ref{eq:calib-o_w}). Together with the
ambiguities expected by design, i.e., arbitrary choice of length units,
object scale and related source–object distance, definition of the
$\phi=0{^\circ}$ orientation within the $x$-$y$ plane and the choice
of origin on the $z$-axis, these ambiguities are a special case of
general projective ambiguities of the form
\begin{equation}
\boldsymbol{P}\vec{x}=\boldsymbol{P}\boldsymbol{H}^{-1}\boldsymbol{H}\vec{x}=(\boldsymbol{P}\boldsymbol{H}^{-1})(\boldsymbol{H}\vec{x})=\tilde{\boldsymbol{P}}\tilde{\vec{x}}\:,\label{eq:general-projective-ambiguities}
\end{equation}
with the homography $\boldsymbol{H}$ being composed of the following
transformations in the present case:
\begin{equation}
\begin{aligned}\boldsymbol{H}^{\phantom{-1}} & =\boldsymbol{R}_{z}(\omega)\,\boldsymbol{T}_{z}(\Delta z)\,\boldsymbol{S}(s_{\mathrm{d}},s_{\mathrm{o}})\,\boldsymbol{D}(\delta,\gamma)\\
\boldsymbol{H}^{-1} & =\boldsymbol{D}^{-1}(\delta,\gamma)\,\boldsymbol{S}(s_{\mathrm{d}}^{-1},s_{\mathrm{o}}^{-1})\,\boldsymbol{T}_{z}(-\Delta z)\,\boldsymbol{R}_{z}(-\omega)\:,
\end{aligned}
\label{eq:homography}
\end{equation}
with\begingroup\addtolength{\jot}{.7em}
\begin{align}
\boldsymbol{R}_{z}(\omega) & =\left[\begin{array}{cccc}
\cos(\omega) & -\sin(\omega) & 0 & 0\\
\sin(\omega) & \phantom{-}\cos(\omega) & 0 & 0\\
0 & 0 & 1 & 0\\
0 & 0 & 0 & 1
\end{array}\right] & \boldsymbol{T}_{z}(\Delta z) & =\left[\begin{array}{cccc}
1 & 0 & 0 & 0\\
0 & 1 & 0 & 0\\
0 & 0 & 1 & \Delta z\\
0 & 0 & 0 & 1
\end{array}\right]\label{eq:rot-trans-homography}\\
\boldsymbol{S}(s_{\mathrm{d}},s_{\mathrm{o}}) & =s_{\mathrm{d}}\left[\begin{array}{cccc}
1 & 0 & 0 & 0\\
0 & 1 & 0 & 0\\
0 & 0 & 1 & 0\\
0 & 0 & 0 & s_{\mathrm{o}}
\end{array}\right]\label{eq:scale-homography}\\
\boldsymbol{D}(\delta,\gamma) & =\left[\begin{array}{cccc}
1 & 0 & 0 & 0\\
0 & 1 & 0 & 0\\
0 & 0 & 1/\text{\ensuremath{\delta}} & 0\\
0 & 0 & \gamma & 1
\end{array}\right] & \boldsymbol{D}^{-1}(\delta,\gamma) & =\left[\begin{array}{cccc}
1 & 0 & 0 & 0\\
0 & 1 & 0 & 0\\
0 & 0 & \text{\ensuremath{\delta}} & 0\\
0 & 0 & -\delta\gamma & 1
\end{array}\right]\:.\label{eq:delta-gamma-homography}
\end{align}
\endgroup Rotation $\boldsymbol{R}_{z}(\omega)$, $z$-translation
$\boldsymbol{T}_{z}(\Delta z)$ and gobal scale $s_{\mathrm{d}}$
correspond to a choice of reference frame and length unit that can
be made at the users convenience. The object scale $s_{\mathrm{o}}$
affects both the object size and the relation between source–axis
and detector–axis distance accordingly and can only be determined
based on prior knowledge on the sample and will therefore be arbitrarily
fixed to $s_{\mathrm{o}}=1$ here. $\boldsymbol{D}(\delta,\gamma)$
describes the initially mentioned independence of Eqs.~\ref{eq:calib-a_h}–\ref{eq:calib-a_w}
and Eqs.~\ref{eq:calib-o_h}–\ref{eq:calib-o_w} and affects the
object and detector geometry respectively and is therefore of particular
interest here. Examples of the ambiguities described by $\boldsymbol{D}$
and $\boldsymbol{S}$ are depicted in Figure~\ref{fig:projective-ambiguity}.

 In contrast to the other transformations, $\boldsymbol{D}(\delta,\gamma)$
can be constrained by drawing upon further knowledge on the imaging
system, whose pixel aspect ratio as well as the angle between detector
rows and columns is commonly known. Anticipating the derivations given
in Sections~\ref{subsec:pixel-aspect-ratio}–\ref{subsec:detector-tilt},
it can be summarized that $\delta$ affects the detector pixel aspect
ratio, while $\gamma$ simultaneously affects the pixel shear, aspect
ratio and the tilt angle of the detector plane. With $\gamma=0$,
$\delta$ may be chosen according to Equation~\ref{eq:aspect-ratio-correction}
in order to enforce a given detector pixel aspect ratio, whereby
$\varepsilon=1$ for many typical detectors. Otherwise, $\gamma$
and $\delta$ need to be determined simultaneously by optimization
of an objective function as defined by Equation~\ref{eq:homography-paramters-objective}
which approaches $0$ for rectangular detector pixels (i.e., $\vec{H}\cdot\vec{V}=0$)
of aspect ratio $\varepsilon$ (i.e., $\bigl\Vert\vec{H}\bigr\Vert/\bigl\Vert\vec{V}\bigr\Vert=\varepsilon$).
In the case of zero detector slant about the rotational axis, Equation~\ref{eq:homography-paramters-objective}
has no unique minimum and therefore many solutions for $\gamma$ and
$\delta$. This degeneracy of auto calibration with respect to detector
tilt has also been described e.g.\ by Smekal et al.\ \cite{Smekal2004}
and can further be related to the general difficulty to precisely
determine detector tilt, which has been reported in context of many
different calibration approaches.

Besides constraints on the detector geometry, knowledge on the sample
such as relative distances between markers (as often used in previous
literature) could be used as well in order to determine the two parameters
defining $\boldsymbol{D}$. This however shall not be further considered
here given the specific focus on calibration based on unknown marker
locations.

\subsection{Relation between Real Space Geometry and Projection Matrices}

With respect to the final objective to conversely transform projection
matrices to real space vectors describing source and detector location
as well as detector row and column orientations (see Figure~\ref{fig:geometry-def}),
first the formulation of projection matrices in terms of real space
geometry vectors will be discussed. The derived relations between
projection matrices and vector geometries will be required both for
the resolution of projective ambiguities and for general real space
interpretations of the calibration results.

\subsubsection{Formulation of Projection Matrices in Terms of Vector Geometry}

To begin with, a vectorial formulation of the rays running between
source and detector shall be given to this end: An euclidean point
$\vec{p}$ somewhere on the connecting line between source $\vec{s}$
and a detector pixel $(\vec{d}+h\vec{H}+v\vec{V})$ characterized
by the detector location $\vec{d}$, the pixel's coordinate $(h,v)$
on the detector matrix and the detector row and column strides $\vec{H}$
and $\vec{V}$ relating the matrix to real space, is described by
the parametric equation

\begin{equation}
\vec{p}(h,v,w)=\vec{s}+w\underbrace{\Bigl((\overbrace{\vec{d}+h\vec{H}+v\vec{V}}^{\mathclap{\substack{\text{position of pixel }(h,v)\\
\text{in 3D space}
}
}})-\vec{s}\Bigr)}_{\text{line orientation}}\label{eq:projection-line}
\end{equation}
with $w$ characterizing the relative position along the line between
source $\vec{s}$ and detector pixel $(\vec{d}+h\vec{H}+v\vec{V})$.

The projection $(h,v)$ of a given point $\vec{p}$ is given by the
inverse of the above relation. By choice of a more convenient representation
of Eq.~\ref{eq:projection-line} introducing the auxiliary variables
$h'$ and $v'$, the inversion can be formulated as follows:
\begin{align}
\vec{p} & =\vec{s}+\overbrace{w\,h}^{\smash[t]{h'}}\vec{H}+\overbrace{\vphantom{h}w\,v}^{\smash[t]{v'}}\vec{V}+w\,(\vec{d}-\vec{s})\label{eq:raycasting-2}\\
\vec{p}-\vec{s} & =\left[\begin{array}{c}
\!\vec{H}\:\Bigl|\:\vec{V}\:\Bigr|\:\vec{d}-\vec{s}\end{array}\right]\left[\begin{array}{c}
\!h\mathrlap{{}'}\!\\
\!v\mathrlap{{}'}\!\\
\!w\!
\end{array}\right]\nonumber \\
\smash[t]{\left[\begin{array}{c}
\!h\mathrlap{{}'}\!\\
\!v\mathrlap{{}'}\!\\
\!w\!
\end{array}\right]} & =\left[\begin{array}{c}
\!\vec{H}\:\Bigl|\:\vec{V}\:\Bigr|\:\vec{d}-\vec{s}\end{array}\right]^{-1}\Bigl[\vec{p}-\vec{s}\,\Bigr]\label{eq:raycasting-inverse-1}\\
\left[\begin{array}{c}
h\\
v
\end{array}\right] & =\frac{1}{w}\left[\begin{array}{c}
h'\\
v'
\end{array}\right]\:.\label{eq:raycasting-inverse-2}
\end{align}
This solution to the problem $(h(\vec{p}),v(\vec{p}))$ can be straight
forwardly reformulated to reproduce the projection matrix formalism:
\begin{equation}
\overbrace{\negthickspace\negthickspace\left[\begin{array}{c}
h\mathrlap{{}'}\\
v\mathrlap{{}'}\\
w
\end{array}\right]\negthickspace\negthickspace}^{\mathclap{\substack{2\mathrm{D}+1\text{ homoge-}\\
\text{neous coordinate}
}
}}\;\:=\Bigl[\,\overbrace{\underbrace{\Bigl[\begin{array}{c}
\!\vec{H}\:\Bigl|\:\vec{V}\:\Bigr|\:\vec{d}-\vec{s}\end{array}\Bigr]^{-1}}_{\boldsymbol{P}_{3\times3}}\:\Bigr|\:\underbrace{\Bigl[\begin{array}{c}
\!\vec{H}\:\Bigl|\:\vec{V}\:\Bigr|\:\vec{d}-\vec{s}\end{array}\Bigr]^{-1}\Bigl[-\vec{s}\,\Bigr]}_{\boldsymbol{P}_{4}}\,}^{3\times4\text{ projection matrix }\text{\ensuremath{\boldsymbol{P}}}}\Bigr]\Bigl[\overbrace{\begin{array}{c}
\vec{p}\\
1
\end{array}}^{\mathclap{\substack{3\mathrm{D}+1\\
\text{homogeneous}\\
\text{coordinate}
}
}}\Bigr]\:,\label{eq:raycasting-to-pmat}
\end{equation}
where the remaining constant ($\vec{s}$) of the projection geometry
has been included within the fourth column ($\boldsymbol{P}_{4}$)
of the projection matrix $\boldsymbol{P}$. By means of the corresponding
fourth component added to the to-be-projected point $\vec{p}$, the
projection matrix formalism exactly reproduces Eq.~\ref{eq:raycasting-inverse-1}.
All constants of the projection geometry are now completely contained
within $\boldsymbol{P}$, and the projection takes the form of a linear
map between homogeneous coordinates. An explicit representation of
$\boldsymbol{P}$ is found by application of Cramer's rule:
\begin{align}
\begin{aligned}\boldsymbol{P} & =\alpha\left[\begin{array}{c}
\phantom{-}(\phantom{\vec{H}}\mathllap{\vec{V}}\times(\vec{d}-\vec{s}))^{T}\\
-(\vec{H}\times(\vec{d}-\vec{s}))^{T}\\
\phantom{-}(\vec{H}\times\mathrlap{\vec{V}}\phantom{(\vec{d}-\vec{s})})^{T}
\end{array}\left|\begin{array}{c}
-(\phantom{\vec{H}}\mathllap{\vec{V}}\times\mathrlap{\vec{d}}\phantom{\vec{V}})\cdot\vec{s}\\
\phantom{-}(\vec{H}\times\mathrlap{\vec{d}}\phantom{\vec{V}})\cdot\vec{s}\\
-(\vec{H}\times\vec{V})\cdot\vec{s}
\end{array}\right.\right]\\
\text{with}\quad\alpha & =\det\left(\Bigl[\begin{array}{c}
\!\vec{H}\:\Bigl|\:\vec{V}\:\Bigr|\:\vec{d}-\vec{s}\end{array}\Bigr]\right)^{-1}=\left((\vec{H}\times\vec{V})\cdot(\vec{d}-\vec{s})\right)^{-1}\:,
\end{aligned}
\label{eq:pmat-explicit}
\end{align}
where the determinant $\alpha$ can, due to the invariance of projections
with respect to the absolute scale of $\boldsymbol{P}$, be dropped
for all practical purposes. It is here explicitly included for consistency
with Eq.~\ref{eq:raycasting-to-pmat}. 

\subsubsection{Conversion of Projection Matrices to Real Space Vectors\label{subsec:pmat-to-realspace}}

Based on the representation given in Eqs.~\ref{eq:raycasting-to-pmat},
the relation
\begin{equation}
\vec{s}=-\boldsymbol{P}_{3\times3}^{-1}\boldsymbol{P}_{4}\label{eq:pmat-focalpoint}
\end{equation}
can be directly inferred, with $\vec{s}$ being the focal point of
the projection. As can be easily verified, Eq.~\ref{eq:pmat-focalpoint}
is invariant with respect to the absolute scale of $\boldsymbol{P}$.

The remaining vectors $\vec{H}$, $\vec{V}$ and $\vec{d}$ are, in
contrast, related to the absolute scale of $\boldsymbol{P}$. The
ambiguity corresponds to the fact that projections in the coordinate
system of the detection screen are invariant under a proportional
change of scale and distance of that screen, as well as under reflection
through the focal point. In order to recover the screen's actual scale
from an arbitrarily normalized matrix $\boldsymbol{P}$, prior knowledge
such as the true pixel pitch $\bigl\Vert\vec{H}\bigr\Vert$ or $\bigl\Vert\vec{V}\bigr\Vert$
and the screen's orientation with respect to $\vec{s}$ needs to be
incorporated. Nevertheless, a valid preliminary set of equivalent
vectors $\vec{H}'$, $\vec{V}'$ and $\vec{d}'$ reproducing a given
projection matrix $\boldsymbol{P}$ can generally be obtained irrespective
of the original system dimensions based on Eqs.~\ref{eq:raycasting-to-pmat}
and \ref{eq:pmat-focalpoint}:
\begin{equation}
\Bigl[\begin{array}{c}
\!\vec{H}'\:\Bigl|\:\vec{V}'\:\Bigr|\:\vec{d}'-\vec{s}\end{array}\Bigr]=\boldsymbol{P}_{3\times3}^{-1}\label{eq:pmat-to-eqivgeom}
\end{equation}
Based on the assumption typical to X-ray imaging that $(\vec{d}-\vec{s})\cdot\vec{s}<0$,
i.e., that the focal point never lies between the projected field
of view and the projection screen, and further assuming the pixel
pitches $\bigl\Vert\vec{H}\bigr\Vert$ and $\bigl\Vert\vec{V}\bigr\Vert$
to be known, Equations \ref{eq:pmat-focalpoint} and \ref{eq:pmat-to-eqivgeom}
may be completed to
\begin{equation}
\begin{aligned}\vec{s} & =-\boldsymbol{P}_{3\times3}^{-1}\boldsymbol{P}_{4}\\
\Bigl[\begin{array}{c}
\!\vec{H}'\:\Bigl|\:\vec{V}'\:\Bigr|\:\vec{d}'-\vec{s}\end{array}\Bigr] & =\phantom{-}\boldsymbol{P}_{3\times3}^{-1}\\
\Bigl[\begin{array}{c}
\!\mathrlap{\vec{H}}\phantom{\vec{H}'}\:\Bigl|\:\mathrlap{\vec{V}}\phantom{\vec{V}'}\:\Bigr|\:\mathrlap{\vec{d}}\phantom{\vec{d'}}-\vec{s}\end{array}\Bigr] & =-\mathrm{sign}((\vec{d'}-\vec{s})\cdot\vec{s})\smash{\sqrt{\frac{\bigl\Vert\vec{H}\bigr\Vert}{\bigl\Vert\vec{H}'\bigr\Vert}\frac{\bigl\Vert\vec{V}\bigr\Vert}{\bigl\Vert\vec{V}'\bigr\Vert}}}\Bigl[\begin{array}{c}
\!\vec{H}'\:\Bigl|\:\vec{V}'\:\Bigr|\:\vec{d}'-\vec{s}\end{array}\Bigr]\\
\vec{d} & =(\vec{d}-\vec{s})+\vec{s}\:,
\end{aligned}
\label{eq:pmat-to-constrainedgeom}
\end{equation}
whereby the geometric mean over both pixel pitches is meaningful in
the presence of noise on $\vec{H}'$ and $\vec{V}'$, as will be the
case when $\boldsymbol{P}$ is actually determined from experimental
data instead of being explicitly constructed. With respect to the
discussion of projective ambiguities (Sections \ref{subsec:ambiguities}
and \ref{subsec:appendix-ambiguities}) it is further insightful to
explicitly state the inverse $\boldsymbol{P}_{3\times3}^{-1}$ using
Cramer's rule:
\begin{equation}
\begin{aligned}\vec{H}\propto\quad\hphantom{(\vec{d}'-\vec{s})}\mathllap{\vec{H}'} & =\hphantom{-}\alpha^{-1}\,(P_{21},P_{22},P_{23})\times(P_{31},P_{32},P_{33})\\
\vec{V}\propto\quad\hphantom{(\vec{d}'-\vec{s})}\mathllap{\vec{V}'} & =-\alpha^{-1}\,(P_{11},P_{12},P_{13})\times(P_{31},P_{32},P_{33})\\
(\vec{d}-\vec{s})\propto\quad(\vec{d}'-\vec{s}) & =\hphantom{-}\alpha^{-1}\,(P_{11},P_{12},P_{13})\times(P_{21},P_{22},P_{23})\\
\alpha & =\det\boldsymbol{P}_{3\times3}=[(P_{11},P_{12},P_{13})\times(P_{21},P_{22},P_{23})]\cdot(P_{31},P_{32},P_{33})\\
\alpha^{-1} & =\det\boldsymbol{P}_{3\times3}^{-1}=(\vec{H}'\times\vec{V}')\cdot(\vec{d}'-\vec{s})\quad(\text{cf. also Eq. }\ref{eq:pmat-explicit})
\end{aligned}
\label{eq:pmat-inversion-explicit}
\end{equation}

\section{Simulation Study}

\begin{figure}
\centering{}\includegraphics[width=0.9\textwidth]{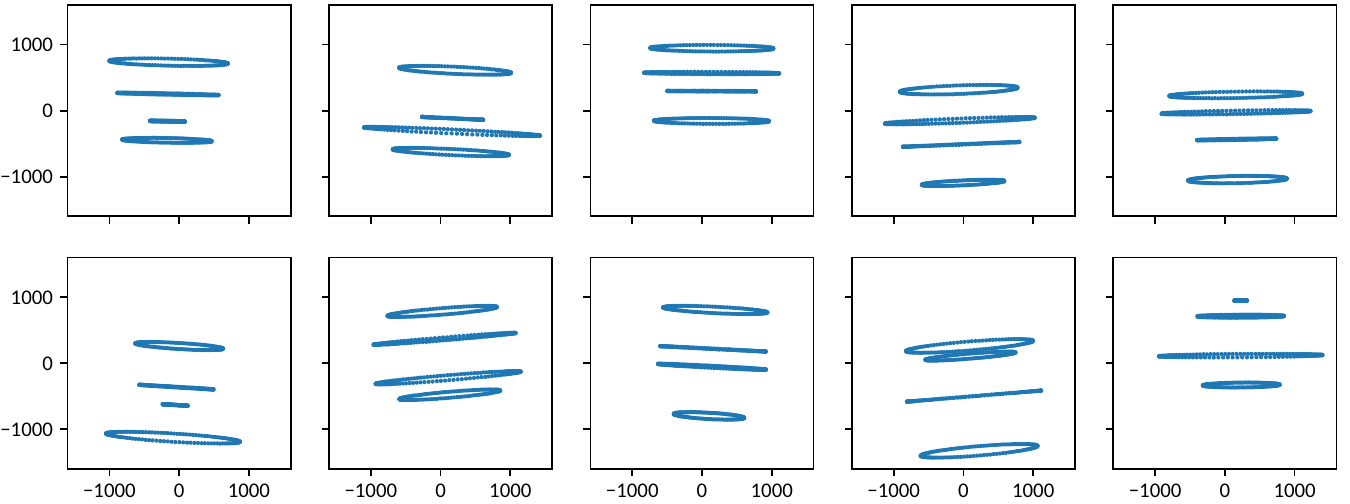}\caption{\label{fig:auto-calib-simulation-examples}Examples of simulated projections
of rotating fiducial markers as used to quantify the auto calibration
precision. Both the projection geometry and the marker placement
are randomly generated. Positions are given in units of detector pixels,
with the origin defined in the detector's center. Auto calibration
is performed both from all shown trajectories (4 markers) as well
as from only the top and bottom ones (2 markers). Cf.\ Fig.~\ref{fig:recon-error-histograms}.}
\end{figure}
The calibration procedure has been tested on simulations of a large
set of randomly generated imaging configurations and fiducial marker
positions in order to obtain information on the average precision
independent of the particular projection geometry or calibration phantom.
The random samples are generated from both a mean imaging configuration
and mean phantom shape with broad variances on the actual positions
and orientations. In units of detector pixels, detectors with roughly
1500 to 3000 pixels width and 1000 to 2000 pixels height at a source–detector
distance of 10000 times the detector pixel size are modeled, yielding
cone angles in the range of about $(12\pm5){^\circ}$. With the rotational
axis virtually placed at the location of the detector, also the sample
units can be meaningfully measured in units of detector pixels. Fiducial
markers are, on average, distributed equidistantly along the $z$-axis
between $-$650 and $+$650 pixels with a mean radius of 800 pixels.
Apart from the source–detector distance, all parameters including
detector shifts and tilts are varied randomly. The marker vertical
positions and radii are varied based on a a normal distribution of
150 and 250 pixels standard deviation respectively. The projection
parameters are varied with uniform distributions. The considered ranges
are $\pm$250 and $\pm$500 pixels for the horizontal and vertical
offsets of the detector center from the optical axis respectively
and $\pm5\text{°}$ for detector tilt, slant and rotation. In order
to avoid the unresolvable projective ambiguity in the case of zero
slant of the detector about the rotational axis, the interval of $[-0.2{^\circ},+0.2{^\circ}]$
has been excluded here. For each configuration, 120 projections in
3° increments about the rotational axis have been calculated and gaussian
noise with a variance of half a pixel was added to the marker projections
to account for imprecisions usually occurring when evaluating actual
projection data of opaque markers. Figure~\ref{fig:auto-calib-simulation-examples}
shows examples of respective simulated projection data. 
\begin{figure}[p]
\noindent \centering{}\includegraphics[width=0.98\textwidth]{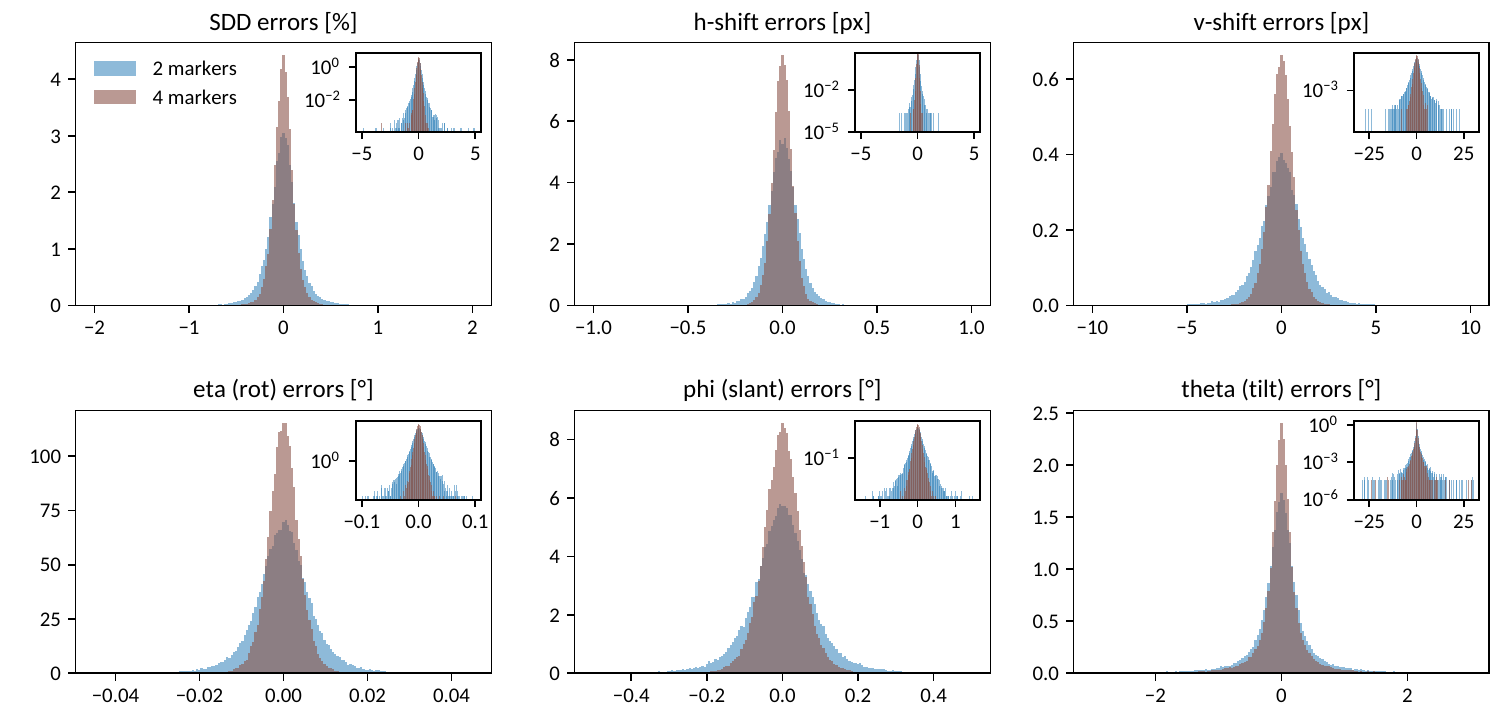}\caption{\label{fig:recon-error-histograms}Normalized histograms (probability
densities) of the reconstruction errors found for projection geometries
reconstructed from noisy (1/2 pixel standard deviation) projections
of circular trajectories. Actual projection geometries and trajectories
are generated randomly. The reconstructions have been performed using
either four (purple) or only two (blue) of the projected trajectories.
The inset graphs visualize the tails of the respective distributions
using a logarithmic scale. Table~\ref{tab:recon-error-intervals}
summarizes the error ranges for a 98\% confidence interval.}
\end{figure}
\begin{table}[p]
\noindent \centering{}%
\begin{tabular}{ccccccc}
\hline 
 & source–detector & horizontal & vertical & detector & detector & detector\tabularnewline
 & distance & detector shift & detector shift & slant ($\varphi$) & rotation ($\eta$) & tilt ($\theta$)\tabularnewline
\hline 
\hline 
4 markers & $\pm0.3\%$ & $\pm0.13$px & $\pm1.7$px & $\pm0.14\text{°}$ & $\pm0.01\text{°}$ & $\pm1.6\text{°}$\tabularnewline
2 markers & $\pm0.5\%$ & $\pm0.22$px & $\pm3.6$px & $\pm0.27\text{°}$ & $\pm0.02\text{°}$ & $\pm2.3\text{°}$\tabularnewline
\hline 
\end{tabular}\caption{\label{tab:recon-error-intervals}Geometry reconstruction errors within
a 98\% confidence interval for reconstructions based on two and four
projected marker trajectories respectively.}
\end{table}
\begin{figure}
\begin{centering}
\includegraphics[width=0.25\textwidth]{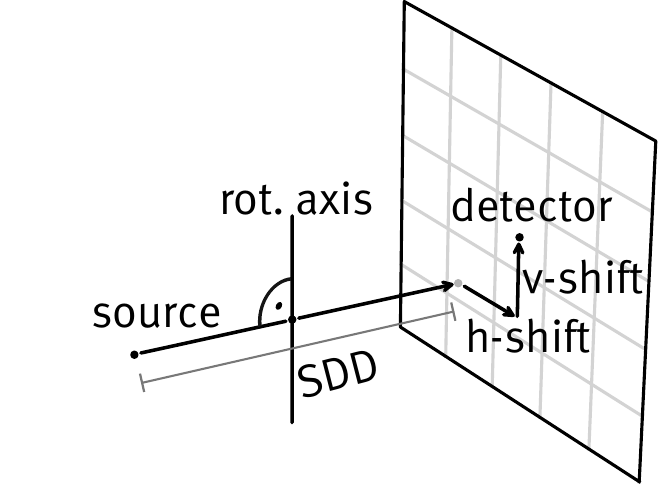}$\quad\quad\quad\quad\quad\quad$\includegraphics[width=0.2\textwidth]{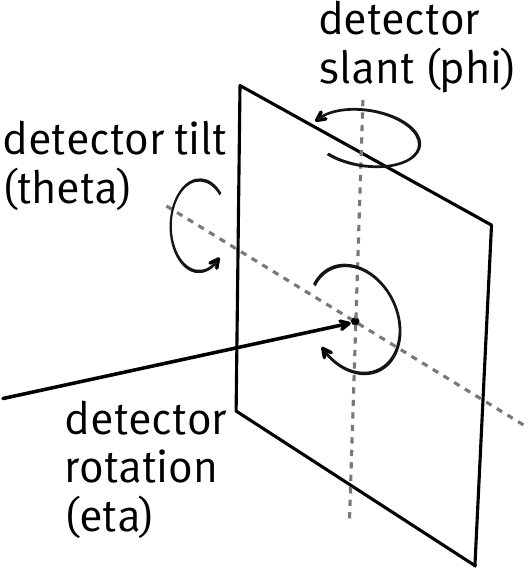}$\quad\quad\quad$
\par\end{centering}
\caption{\label{fig:simulation-geometry-sketches}Sketches depicting the employed
geometry parametrization. The source–detector distance SDD is measured
parallel to the source position vector $\vec{s}$, horizontal ($h$)
and vectical ($v$) detector shifts are measured in units of the row
and column vectors $\vec{H}$ and $\vec{V}$. The detector orientation
is characterized by the angles $\varphi$ and $\theta$ describing
the orientation of its normal with respect to the source orientation
$\vec{s}$ as well as the angle $\varphi$ describing the in-plane
rotation of the detector about its normal.}
\end{figure}

Calibration was performed based on the solution strategy for Eqs.~\ref{eq:calib-a_w}–\ref{eq:calib-phi_w}
derived in Section~\ref{subsec:iterative-pmat-recon}. More specifically,
each projected trajectory is first reduced to its sinusoid parameters
by means of Algorithm~\ref{alg:pmat-recon-sinusoids}. Based on these
observables, a self consistent solution for markers and projection
matrix is found by means of Algorithm~\ref{alg:iterative-pmat-recon}.
The projective ambiguities are resolved by means of Eq. \ref{eq:general-projective-ambiguities},
\ref{eq:delta-gamma-homography} and \ref{eq:homography-paramters-objective}
constraining the detector rows and columns to be orthogonal and the
pixel aspect ratio to be 1. The resulting projection matrix is transformed
into real space geometry vectors describing relative source and detector
position as well as row and column orientation by means of Eq.~\ref{eq:pmat-to-constrainedgeom}
and the known detector pixel pitch. In order to compare this result
to the original (randomly generated) imaging geometry, the reconstructed
geometry is finally rotated and shifted about and along the $z$ axis
respectively such that the source comes to line on the $y$ axis.

Figure~\ref{fig:recon-error-histograms} shows the distribution of
errors found for $10^{6}$ random realizations of the just described
experiment. Table~\ref{tab:recon-error-intervals} summarizes the
error ranges corresponding to a 98\% confidence interval, i.e.\ in
98\% of the cases, the true values will lie within the listed intervals
about the reconstructed values. Figure~\ref{fig:simulation-geometry-sketches}
sketches the chosen geometry parametrization.

Particularly the distribution of errors on the detector tilt towards
the rotational axis exhibits long tails including rare (as seldom
as one in a million) yet extreme deviations (up to 25°) from the true
value. As the detector tilt is part of the projective ambiguity in
the solution of Equations \ref{eq:calib-a_h}–\ref{eq:calib-phi_w}
as detailed in Section~\ref{subsec:ambiguities}, it has to be inferred
by enforcing a known detector pixel geometry. As the latter is implicitly
assumed to be unaffected by noise, any actual noise will translate
onto the remaining parameters of the imaging geometry, causing large
uncertainties on the inferred tilt. Further, as the determined homography
inversely applies to the reconstructed sample, errors on the detector
tilt will come along with errors on the $z$ scale and consequently
on the aspect ratio of the reconstructed samples. 

\section{Discussion}

An auto-calibration method for cone beam tomography systems has been
derived from the projection matrix formulation of the perspective
projection of rotating fiducial markers. The representation in cylinder
coordinates directly reveals both a fractions-of-sinusoids model describing
the observable projections as well as linear relations of this model's
parameters to both the unknown projection matrix and the parameters
of the circular marker trajectories. (In the case of known trajectory
parameters, the system of equations could at this point be directly
solved for the unknown projection matrix analog to other projection
matrix calibration methods based on known objects \cite{RougeeTrousset1993,StrobelWiesent2003,XLiBLiu2010}.)
An iterative scheme is proposed that alternatingly solves the linear
system of equations with respect to the projection matrix and the
trajectories until a self consistent solution is obtained, starting
with initial approximations for the trajectory parameters. A simple
weighting heuristic accounts for the adequate consideration of redundant
equations. The ambiguities in the self consistent solution are formalized
to a sparse homography matrix, which is then constrained based on
knowledge of the detector pixel geometry. Finally, a transformation
of projection matrices into real space vectors describing source and
detector position as well as detector row and column orientations
is given. Other geometry descriptions can be derived from either the
projection matrices or the real space vectors using common techniques
that have not been further detailed here. 

The forward model reveals that the projected circular trajectories
are completely described by a total of 8 parameters (which can be
identified with those that were previously also used by Smekal et
al.\ \cite{Smekal2004}). In contrast to the ellipse description
of projected circles (generally using 5 parameters) that has been
used by many authors \cite{Noo2000,Cho2005,StrubelNoo2005,KYangBoone2006,JohnstonBadea2008,RobertMainprize2009,DWuYXiao2011,GrossSchwanecke2012,XuTsui2013},
not only the shape, but also the projection angle ($\phi$) dependence
is captured correctly in this description. This assumably is the underlying
reason why degeneracy of ellipses, which is a major issue for the
respective methods based on ellipse paramters, is not a particularly
special case in this representation – all 8 parameters are still defined
also in the case of an edge-on projection within the plane of rotation
of a circular orbit. The forward model itself almost directly exposes
the solution to the calibration problem in form of the system of equations
relating the sinusoid parameters to the contained system parameters.
The employed forward model is further clearly separated into orthographic
projections (numerators) and perspective scaling (denominator), allowing
for direct interpretations of the parameters. The denominator or perspective
scaling component is associated with the third row of the projection
matrix (cf.\ Eqs.\ \ref{eq:projection_homogeneus_form}–\ref{eq:projection_perspective_v}),
which in turn describes, as is explicitly shown in Section~\ref{subsec:pmat-to-realspace},
the detector normal and focal distance. This clearly highlights that
the ability to determine the orientation of the detector normal (i.e.,
to determine detector slant and tilt) without knowledge of the sample
dimensions is ultimately founded in the specific perspective scaling
effects associated with the detector orientation. Conversely, this
ability becomes restricted in the case of ambiguous scaling effects:
in the case of zero detector slant (i.e., the rotational axis intersects
the orthogonal connecting line between detector and source), scaling
effects due to a potentially tilted detector become indiscernible
from an actually distorted sample, and, as has been reported previously
by several authors \cite{Rizo1994,Beque2003,WangTsui2007,SawallKachelriess2012},
additional knowledge or assumptions will thus be required (usually,
the assumption of zero detector tilt will be adequate given its typically
small effect \cite{Noo2000,Smekal2004,KYangBoone2006}). In the present
method, this ambiguity manifests itself in a degenerate minimum of
the objective function used to determine the homography parameters
as discussed in Section~\ref{subsec:ambiguities}.

Regarding the positioning or selection of fiducial markers within
the field of view, although no explicit analyses have been shown,
several general arguments can be made. First of all, markers obviously
must not be located on the rotational axis, i.e., the radius of its
circular trajectory must not be zero. The larger the radius relative
to the detector grid, the smaller the relative errors in the determination
of the projected positions. Larger radii further imply larger covered
cone angles and thus more pronounced perspective scaling effects,
which have been identified to be essential to auto-calibration from
unknown samples. Analogously, although edge-on projections are not
a fundamental issue here, trajectories further away from the cone
center plane will exhibit stronger perspective effects and are therefore
expected to be favorable. This is also consistent with the observations
by Bequé et al. \cite{Beque2005} in context of their least squares
optimization approach. Also, placement of markers only within one
half of the cone does not constitute a special case, in contrast to
ellipse based methods as introduced by Noo et al. \cite{Noo2000}.
In order to account for the original assumption of the present work
that markers are apriori unknown, the presented simulation study intentionally
addresses a wide range of imaginable marker placements and projection
geometries, without explicitly investigating potentially favorable
configurations.

One reason for the simplicity of the present calibration approach
lies in the additional degree of freedom of oblique detector grids
that is implicit in the projection matrix formalism. The relation
between detector tilt and detector grid geometry through the homography
parameters $\gamma$ and $\delta$ (cf.\ Sections \ref{subsec:ambiguities}
and \ref{subsec:appendix-ambiguities}) on the one hand allows to
shift the determination of tilt into a downstream postprocessing procedure,
and on the other hand provides another view on the previously reported
imprecision immanent to the determination of detector tilt \cite{Smekal2004,GrossSchwanecke2012},
which could also be observed in the present results. The imprecision
in the determination of tilt is in fact an imprecision in the determination
of the correct homography transformation (affecting both the projection
matrix and the object coordinate system) based on constraints on the
detector geometry instead of constraints on the sample geometry. As
even considerable tilts in the range of degrees translate to rather
moderate amounts of detector non-orthogonality, the noise susceptibility
for the converse inference of tilt by means of constraining detector
shear is very high. In particular the long tails of the error distribution
found in the present simulations (cf.\ Fig.~\ref{fig:recon-error-histograms}
and Table~\ref{tab:recon-error-intervals}) evidence a very high
uncertainty in the determination of tilt that often ranges within
the order of magnitude of its actual value. The plain assumption of
zero tilt is thus, as has been concluded by others previously \cite{Smekal2004,GrossSchwanecke2012},
often well within the error margin. In contrast to previous work though,
which by design also constrained the detector geometry, consistency
won't be affected here, as actual effects due to tilt will still be
accounted for by means of an artificially oblique detector geometry.

The remaining homography parameters regarding the choice of origin
and scale of the coordinate system have not been explicitly treated.
They are straight-forwardly chosen based on the real space geometry
description extracted from the reconstructed projection matrix (cf.
Section~\ref{subsec:pmat-to-realspace}) in order to e.g.\ align
the source position or the detector normal with the $x$-$z$ or $y$-$z$
plane and scale and shift source and detector within the available
degrees of freedom such that a user defined tomographic reconstruction
field of view optimally fits the projection cone.

\section{Conclusion}

A projection matrix based approach to the auto-calibration of the
projection geometry of cone beam computed tomography systems from
apriori unknown fiducial markers moving along circular trajectories
has been proposed. In contrast to previous literature, the problem
of calibration has been decoupled from the particular choice of real
space geometry parametrization, allowing for a very simple representation
of the core problem without having to fall back to generic optimization
approaches. The link to classic, more intuitive geometry representations
is provided by means of explicit conversion formulas of projection
matrices to real space position and orientation vectors. The formulation
of ambiguities in terms of homography transformations further reveals
the relation to methods based on known samples, which essentially
differ in the particular prior knowledge used to constrain the solution
space. The proposed scheme for the self consistent solution of the
derived system of equations both with respect to the unknown projection
and sample parameters has been tested on a large variety of simulated
noisy projection data in order to assess the average precision independent
of the particular instances of marker placement, noise and geometric
configuration of the projection system. A high degree of versatility
is achieved by avoiding the necessity of prior estimates on the system
or sample parameters which is commonly required for methods based
on the optimization of a cost function. Last but not least, the present
formulation provides another, hopefully more intuitive view, on the
auto-calibration problem, allowing further insights into the fundamental
working principle and limitations of auto-calibration of cone beam
computed tomography systems from unknown samples.

\section{Acknowledgments}

The author thanks Prof. Dr. Randolf Hanke and Dr. habil. Simon Zabler
for facilitating the present work. Dr. Christian Fella, Dr. Kilian
Dremel and Dominik Müller are acknowledged for providing experimental
data used to test the presented methods during the stages of development.
Funding is acknowledged from the Bavarian State Ministry of Economic
Affairs, Infrastructure, Transport and Technology which supported
the project group ”Nano-CT Systems for Material Characterization'',
the European Horizon 2020 project no.\ 814485 (LEE-BED), and the
German Federal Ministry of Education and Research grant 05E19AN1 supporting
the BM18 beamline at the European Synchrotron Radiation Facility ESRF.

\section{Appendix}

\subsection{Projective ambiguities\label{subsec:appendix-ambiguities}}

\subsubsection{Detector pixel aspect ratio\label{subsec:pixel-aspect-ratio}}

When considering the relations $r_{i}'P_{m\mathrm{a}}=a_{im}$ (Eqs.\
\ref{eq:calib-a_h}–\ref{eq:calib-a_w}) and assuming some consistent
solutions for $r_{i}'$ and $P_{m\mathrm{a}}$ have already been found,
it is easy to see that these solutions may be scaled by an arbitrary
factor $\alpha\neq0$:
\[
r_{i}'P_{m\mathrm{a}}=r_{i}'\frac{\alpha}{\alpha}P_{m\mathrm{a}}=(\alpha r_{i}')(\frac{1}{\alpha}P_{m\mathrm{a}})=\tilde{r}_{i}'\tilde{P}_{m\mathrm{a}}=a_{im}\quad.
\]
The same is true for $z_{i}'$ and $P_{m3}$ (Eqs.\ \ref{eq:calib-o_h}–\ref{eq:calib-o_w}):
\[
z_{i}'P_{m3}=z_{i}'\frac{\beta}{\beta}P_{m3}=(\beta z_{i}')(\frac{1}{\beta}P_{m3})=\tilde{z}_{i}'\tilde{P}_{m3}=o_{im}
\]
with an independent scaling parameter $\beta\neq0$. While the freedom
to choose an arbitrary overall scale corresponds to the unknown absolute
size of the imaged phantom, the freedom to choose \emph{independent}
scales for the $x$-$y$ and the $z$ dimensions or equivalently for
$r_{i}'$ and $z_{i}'$ or $P_{m\mathrm{a}}$ and $P_{m3}$ corresponds
to the disregarded pixel pitches of the detector, i.e.\ the calibration
equations can equally be satisfied by a detector with asymmetric pixels
and a correspondingly squeezed or stretched object.

Conversely, this ambiguity corresponding to the homography parameter
$\delta$ can be resolved by choosing the relative scale $\alpha/\beta=\delta_{\varepsilon}$
such that the detector encoded in $\boldsymbol{P}$ actually features
the correct pixel aspect ratio (denoted by $\varepsilon$) for the
given hardware. As derived in Section~\ref{subsec:pmat-to-realspace},
the vector products $(P_{11},P_{12},P_{13})\times(P_{31},P_{32},P_{33})$
and $(P_{21},P_{22},P_{23})\times(P_{31},P_{32},P_{33})$ are proportional
to the detector row and column vectors $\vec{H}$ and $\vec{V}$ (Fig.~\ref{fig:geometry-def},
Eq.\ \ref{eq:pmat-inversion-explicit}). Given a known pixel aspect
ratio $\varepsilon\overset{!}{=}\bigl\Vert\vec{H}\bigr\Vert/\bigl\Vert\vec{V}\bigr\Vert$,
$\delta_{\varepsilon}$ may therefore be defined as the solution to
\begin{align*}
\varepsilon\bigl\Vert\vec{V}(\delta_{\varepsilon})\bigr\Vert & \overset{!}{=}\bigl\Vert\vec{H}(\delta_{\varepsilon})\bigr\Vert\\
\varepsilon\left\Vert (P_{11},P_{12},\delta_{\varepsilon}P_{13})\times(P_{31},P_{32},\delta_{\varepsilon}P_{33})\right\Vert  & =\left\Vert (P_{21},P_{22},\delta_{\varepsilon}P_{23})\times(P_{31},P_{32},\delta_{\varepsilon}P_{33})\right\Vert \quad,
\end{align*}
i.e.\
\begin{equation}
\begin{aligned}\delta_{\epsilon} & =\sqrt{\frac{(P_{21}P_{32}-P_{22}P_{31})^{2}-\varepsilon^{2}(P_{11}P_{32}-P_{12}P_{31})^{2}}{\splitfrac{\varepsilon^{2}(P_{12}P_{33}-P_{13}P_{32})^{2}+\varepsilon^{2}(P_{13}P_{31}-P_{11}P_{33})^{2}}{-(P_{22}P_{33}-P_{23}P_{32})^{2}-(P_{23}P_{31}-P_{21}P_{33})^{2}}}}\\
 & =\sqrt{\frac{H_{z}^{2}-\varepsilon^{2}V_{z}^{2}}{\varepsilon^{2}(V_{x}^{2}+V_{y}^{2})-(H_{x}^{2}+H_{y}^{2})}}\quad,
\end{aligned}
\label{eq:aspect-ratio-correction}
\end{equation}
where the $H_{x},H_{y},H_{z}$ and $V_{x},V_{y},V_{z}$ components
refer here to those prior to the correction by the derived scaling
factor $\delta_{\varepsilon}$ to be applied to $P_{m3}$ (and inversely
to $z_{i}'$) in order to make $\boldsymbol{P}$ (and in consequence
also the corresponding euclidean vectors $\vec{s}$, $\vec{d}$, $\vec{H}$
and $\vec{V}$) consistent with the known detector pixel aspect ratio
$\varepsilon$ (which commonly will be $\varepsilon=1$).

\subsubsection{Detector tilt and shear\label{subsec:detector-tilt}}

Also the relations $P_{m3}z_{i}'+P_{m4}w_{i}=o_{im}$ (Eqs.~\ref{eq:calib-o_h}–\ref{eq:calib-o_w})
may similarly be satisfied by transformed $\tilde{P}_{m3}$, $\tilde{z}_{i}'$
and $\tilde{w}_{i}$, now including also $\gamma$:
\[
P_{m3}z_{i}'+P_{m4}w_{i}=\delta(P_{m3}-\gamma P_{m4})(\frac{1}{\delta}z_{i}')+P_{m4}(w_{i}+\gamma z_{i}')=\tilde{P}_{m3}\tilde{z}_{i}'+P_{m4}\tilde{w}_{i}=o_{im}\quad.
\]
The role of $\text{\ensuremath{\delta}}$ has just been discussed,
wherefore $\delta$ will be assumed to equal $1$ for now. The parameter
$\gamma$ relating $w_{i}$ and $P_{m3}$ will in effect control the
detector tilt and shear as will be explained in the following.

Equation~\ref{eq:calib-o_w} ($P_{33}z_{i}'+P_{34}w_{i}=o_{i\mathrm{w}}=1$)
concerning the mean $o_{i\mathrm{w}}=1$ of the projection equations'
denominators (cf.\ Eqs.\ \ref{eq:sinusoid-fw-model_h} and \ref{eq:sinusoid-fw-model_v})
reveals that for $w_{i}=\mathrm{const.}$, $P_{33}$ must equal $0$.
Vice versa, $w_{i}$ varying among several imaged trajectories $i$
(i.e.\ being $z_{i}'$-dependent) implies $P_{33}\neq0$. Given $\tilde{w}_{i}=w_{i}+\gamma z_{i}'$
and $\tilde{P}_{33}=P_{33}-\gamma P_{34}$, it can therefore be concluded
that both the relevance of the homogeneous coordinates' scaling components
$w_{i}$ and the role of the homography parameter $\gamma$ are directly
related to the $P_{33}$ component of the projection matrix. Section~\ref{subsec:pmat-to-realspace}
shows that the vector $(P_{31},P_{32},P_{33})$ in the last row of
$\boldsymbol{P}$ corresponds to the cross product $\vec{H}\times\vec{V}$
of the detector row and column orientations and is therefore normal
to the detector plane. The $z$ component $P_{33}$ is thus directly
related to the detector tilt towards the rotational axis, which was
defined to coincide with the $z$-axis. In consequence, $\gamma$
controls the detector tilt encoded in $\tilde{\boldsymbol{P}}$. As
changes to $P_{m3}$ have more general implications on the cross products
of $(\vec{d}-\vec{s})$, $\vec{H}$ and $\vec{V}$ encoded in the
first three columns of $\boldsymbol{P}$, $\gamma$ will also influence
the angle between $\vec{H}$ and $\vec{V}$. As this is commonly a
known property of the detector (typically, $\vec{H}\cdot\vec{V}=0$),
it can be used to constrain $\gamma$ and consequently to determine
the detector tilt towards the $z$-axis. As $\gamma$ will as well
affect the norms of $\vec{H}$ and $\vec{V}$, i.e.\ the pixel aspect
ratio, $\gamma$ usually needs to be determined simultaneously with
$\delta$ using a suiting objective function constraining both the
pixel aspect ratio and orthogonality:
\begin{equation}
\gamma,\delta_{\varepsilon}=\underset{\gamma,\delta}{\mathrm{argmin}}\left(\left(\frac{\Bigl\Vert\vec{H}(\gamma,\delta)\Bigr\Vert-\varepsilon\Bigl\Vert\vec{V}(\gamma,\delta)\Bigr\Vert}{\Bigl\Vert\vec{H}(\gamma,\delta)\Bigr\Vert+\varepsilon\Bigl\Vert\vec{V}(\gamma,\delta)\Bigr\Vert}\right)^{2}+\left(\frac{\vec{H}(\gamma,\delta)\cdot\vec{V}(\gamma,\delta)}{\Bigl\Vert\vec{H}(\gamma,\delta)\Bigr\Vert\,\Bigl\Vert\vec{V}(\gamma,\delta)\Bigr\Vert}\right)^{2}\right)\label{eq:homography-paramters-objective}
\end{equation}
with $\varepsilon$ denoting the pixel aspect ratio and assuming that
the detector rows and columns are expected to be orthogonal. In case
of actually non-orthogonal detectors such as hexagonal pixel arrangements,
the objective function may be adjusted accordingly to favor the respective
expected shear angle.

The detector tilt has been identified by many authors to have the
smallest influence on the observable projections and therefore can,
in the presence of noise, only be determined with very little precision.
In particular in the context of calibration based on unknown phantoms,
the detector is therefore often fixed to be parallel to the rotational
axis. For the method presented here, this is equivalent to fixing
$w_{i}$ to $1$ and consequently $P_{33}$ to $0$, either by choice
of $\gamma$ or simply by directly choosing the $w_{i}$-related relaxation
parameter $\lambda_{w}=0$ within the iterative reconstruction of
$\boldsymbol{P}$ (cf.\ Section~\ref{subsec:iterative-pmat-recon}).
Actual detector tilts will then manifest themselves in a slight amount
of artificial shear (slightly non-orthogonal detector rows and columns)
in the determined projection geometry. As this corresponds to a valid
projective homography, resulting tomographic reconstructions using
so constrained geometries will be transformed by the corresponding
inverse homography. In contrast to the case of both constrained tilt
\emph{and} pixel geometry, actual artifacts due to geometric inconsistencies
are avoided.

In accordance with the findings by Smekal et al.\ \cite{Smekal2004},
the tilt cannot be uniquely determined when the detector is not slanted
about the rotational axis. This remaining ambiguity manifests itself
in a non-unique minimum of the objective function in Eq.~\ref{eq:homography-paramters-objective}
in the case of zero detector slant.

\subsection{Iterative projection matrix reconstruction\label{subsec:iterative-pmat-recon}}

\begin{algorithm}
\algvspace\algvspace
\begin{spacing}{1.4}
\begin{algorithmic}[H]
\State $\negmedspace\negmedspace\left.\def\arraystretch{.7}\begin{array}{lll}a_{i\mathrm{h}} & \phi_{i0\mathrm{h}} & o_{i\mathrm{h}}\\a_{i\mathrm{v}} & \phi_{i0\mathrm{v}} & o_{i\mathrm{v}}\\a_{i\mathrm{w}} & \phi_{i0\mathrm{w}} & o_{i\mathrm{w}}\end{array}\right\} \gets \;\mathrm{Algorithm}~\ref{alg:pmat-recon-sinusoids}~(h_i(\phi),v_i(\phi))\;\forall\,i$ \Comment{\negmedspace\negmedspace$\def\arraystretch{0.7}\begin{array}{l}\text{Sinusoid parameters extracted}\\\text{from projection data }\end{array}\negmedspace\negmedspace\negmedspace$ }\vspace{.5em}
\State $r_{i}'^{(0)} \gets a_{i\mathrm{h}}$ \Comment{initialize radii with horizontal trajectory amplitudes}
\State $z_{i}'^{(0)} \gets o_{i\mathrm{v}}$ \Comment{initialize $z$-coordinates with vertical means}
\State $P_{m\mathrm{a}}^{(0)} \gets \vphantom{\smash[b]{\Bigg(}} \frac{\sum_{i}r_{i}'^{(0)}a_{im}}{\sum_{i}r_{i}'^{(0)\,2}}$\Comment{initialize with (weighted) mean amplitudes}
\State $P_{m3}^{(0)} \gets \vphantom{\Bigg(} \frac{\sum_{i}(z_{i}'^{(0)}-\overline{z'}^{(0)})(o_{im}-\overline{o}_{m})}{\sum_{i}(z_{i}'^{(0)}-\overline{z'}^{(0)})^{2}}$\Comment{$P_{m3}^{(0)},P_{m4}^{(0)} = \underset{P_{m3},P_{m4}}{\mathrm{argmin}}\left(P_{m3}\,z_{i}'^{(0)}+P_{m4}-o_{im}\right)$}
\State $P_{m4}^{(0)} \gets \vphantom{\smash[t]{\Bigg(}} \overline{o}_{m}-P_{m3}^{(0)}\,\overline{z'}^{(0)}$\Comment{$\mathrlap{\text{with}\quad\overline{z'}^{(0)}=\frac{1}{N}\sum_{i}z_{i}'^{(0)};\quad \overline{o}_{m}=\frac{1}{N}\sum_{i}o_{im}}\hphantom{P_{m3}^{(0)},P_{m4}^{(0)} = \underset{P_{m3},P_{m4}}{\mathrm{argmin}}\left(P_{m3}\,z_{i}'^{(0)}+P_{m4}-o_{im}\right)}$}
\For{$k=0\,..\,N_\mathrm{iter}-1$} with $\lambda\in\;]0;1]$\Comment{iteratively enforce Eqs.~\ref{eq:calib-a_h}--\ref{eq:calib-o_w}}
  \State $r_{i}'^{(k+1)} \gets \vphantom{\Bigg(} (1-\lambda)r_{i}'^{(k)}+\lambda\,\frac{\sum_{m}P_{m\mathrm{a}}^{(k)}a_{im}}{\sum_{m}P_{m\mathrm{a}}^{(k)\,2}}$
  \State $z_{i}'^{(k+1)} \gets (1-\lambda)z_{i}'^{(k)}+\lambda\frac{\sum_{m}P_{m3}^{(k)}(o_{im}-P_{m4}^{(k)})}{\sum_{m}P_{m3}^{(k)\,2}}$
  \State $P_{m\mathrm{a}}^{(k+1)} \gets \vphantom{\Bigg(}(1-\lambda)P_{m\mathrm{a}}^{(k)}+\lambda\,\frac{\sum_{i}r_{i}'^{(k)}a_{im}}{\sum_{i}r_{i}'^{(k)\,2}}$
  \State $P_{m3}^{(k+1)} \gets (1-\lambda)P_{m3}^{(k)}+\lambda\,\frac{\sum_{i}(z_{i}'^{(k)}-\overline{z'}^{(k)})(o_{im}-\overline{o}_{m})}{\sum_{i}(z_{i}'^{(k)}-\overline{z'}^{(k)})^{2}}$\Comment{$\underset{P_{m3},P_{m4}}{\mathrm{argmin}}\left(P_{m3}\,z_{i}'^{(k)}+P_{m4}-o_{im}\right)$}
  \State $P_{m4}^{(k+1)} \gets \vphantom{\Bigg(}(1-\lambda)P_{m4}^{(k)}+\lambda\,\big(\overline{o}_{m}-P_{m3}^{(k)}\,\overline{z'}^{(k)}\big)$\Comment{$\mathrlap{\overline{z'}^{(k)}\!\!=\!\frac{1}{N}\!\sum_{i}z_{i}'^{(k)};\; \overline{o}_{m}\!=\!\frac{1}{N}\!\sum_{i}o_{im}}\hphantom{\underset{P_{m3},P_{m4}}{\mathrm{argmin}}\left(P_{m3}\,z_{i}'^{(k)}+P_{m4}-o_{im}\right)}$}
\EndFor
\State $\phi_{\mathrm{h}} \gets 0$
\State $\phi_{\mathrm{v}} \gets \arg\left(\sum_{i}(e^{\boldsymbol{i}(\phi_{i0\mathrm{v}}-\phi_{i0\mathrm{h}})}a_{i\mathrm{v}}^{2}a_{i\mathrm{h}}^{2})\right)$
\State $\phi_{\mathrm{w}} \gets \arg\left(\sum_{i}(e^{\boldsymbol{i}(\phi_{i0\mathrm{w}}-\phi_{i0\mathrm{h}})}a_{i\mathrm{w}}^{2}a_{i\mathrm{h}}^{2})\right)$

\State $\phi_{i0} \gets \arg\left(\sum_{m}(e^{\boldsymbol{i}(\phi_{i0m}-\phi_{m})}a_{im}^{2})\right)$
\\$\boldsymbol{P}\gets
\left[\begin{array}{cccc} 
 -P_{1\mathrm{a}}\sin(\phi_{\mathrm{h}}) & P_{1\mathrm{a}}\cos(\phi_{\mathrm{h}}) & P_{13} & P_{14} \\
 -P_{2\mathrm{a}}\sin(\phi_{\mathrm{v}}) & P_{2\mathrm{a}}\cos(\phi_{\mathrm{v}}) & P_{23} & P_{14} \\
 -P_{3\mathrm{a}}\sin(\phi_{\mathrm{w}}) & P_{3\mathrm{a}}\cos(\phi_{\mathrm{w}}) & P_{33} & P_{14}
\end{array}\right];\quad\quad
\boldsymbol{x}_i \gets \left[\begin{array}{c} 
 \hphantom{-}r^\prime_i \cos(\phi_{i0}) \\
 -r^\prime_i \sin(\phi_{i0}) \\
 z^\prime_i \\
 1
\end{array}\right]$\vspace{1em}
\State $\boldsymbol{P} \gets \boldsymbol{P}\boldsymbol{H}^{-1};\quad \boldsymbol{x}_i \gets \boldsymbol{H}\boldsymbol{x}_i$ \Comment{resolve projective ambiguities with additional contraints, \\\vspace{-0.285\baselineskip}\hspace*{\fill}cf.\ Sections \ref{subsec:ambiguities} and \ref{subsec:appendix-ambiguities}, especially Eqs.\ \ref{eq:general-projective-ambiguities}, \ref{eq:delta-gamma-homography}, \ref{eq:aspect-ratio-correction}, \ref{eq:homography-paramters-objective}}
\State $\vec{s}, \vec{d}, \vec{H}, \vec{V} \gets \text{Eq.~\ref{eq:pmat-to-constrainedgeom}}$ \Comment{Transform to real space representation and potentially apply\\\vspace{-0.285\baselineskip}\hspace*{\fill}further application-specific coordinate transformations}
\end{algorithmic}
\end{spacing}
\algvspace 

\caption[Self-consistent reconstruction of projection matrix and marker coordinates
from projections of circular marker trajectories.]{\label{alg:iterative-pmat-recon}Self-consistent reconstruction of
projection matrix and marker coordinates from projections of circular
marker trajectories (solving Eqs.~\ref{eq:calib-a_h}–\ref{eq:calib-phi_w}).
The homogeneous scaling components $w_{i}$ are, without loss of generality,
constrained to $1$ (thereby constraining the detector tilt angle
towards the rotational axis to $\theta=0$). Differing values of tilt
and scaling of the marker trajectories can be recovered subsequently
based on projective homographies as outlined in Sections \ref{subsec:ambiguities}
and \ref{subsec:appendix-ambiguities}.}
\end{algorithm}
\begin{algorithm}
\algvspace
\begin{spacing}{1.2}
\begin{algorithmic}
\State $\mathrlap{\phi_{j}}\hphantom{h_{\mathrm{cc}_{2}}}=2\pi/N\,j$
\State $\mathrlap{h_j}\hphantom{h_{\mathrm{cc}_{2}}}=h(\phi_j)$
\State $\mathrlap{v_j}\hphantom{h_{\mathrm{cc}_{2}}}=v(\phi_j)$
\end{algorithmic}
\begin{multicols}{2}
\begin{algorithmic}
\State $\mathrlap{\overline{h}}\hphantom{h_{\mathrm{cc}_{2}}}=\frac{1}{N}\sum_{j=0}^{N-1}h_{j}$
\State $\mathrlap{h_{\mathrm{s}}}\hphantom{h_{\mathrm{cc}_{2}}}=\frac{1}{N}\sum_{j=0}^{N-1}h_{j}\sin(\phi_{j})$
\State $\mathrlap{h_{\mathrm{c}}}\hphantom{h_{\mathrm{cc}_{2}}}=\frac{1}{N}\sum_{j=0}^{N-1}h_{j}\cos(\phi_{j})$ 
\State $\mathrlap{h_{\mathrm{ss}}}\hphantom{h_{\mathrm{cc}_{2}}}=\frac{1}{N}\sum_{j=0}^{N-1}h_{j}\sin^{2}(\phi_{j})$
\State $\mathrlap{h_{\mathrm{sc}}}\hphantom{h_{\mathrm{cc}_{2}}}=\frac{1}{N}\sum_{j=0}^{N-1}h_{j}\sin(\phi_{j})\cos(\phi_{j})$
\State $\mathrlap{h_{\mathrm{cc}}}\hphantom{h_{\mathrm{cc}_{2}}}=\frac{1}{N}\sum_{j=0}^{N-1}h_{j}\cos^{2}(\phi_{j})$

\State $\mathrlap{h_{\mathrm{s_{2}}}}\hphantom{h_{\mathrm{cc}_{2}}}=\frac{1}{N}\sum_{j=0}^{N-1}h_{j}\sin(2\phi_{j})$
\State $\mathrlap{h_{\mathrm{c_{2}}}}\hphantom{h_{\mathrm{cc}_{2}}}=\frac{1}{N}\sum_{j=0}^{N-1}h_{j}\cos(2\phi_{j})$
\State $\mathrlap{h_{\mathrm{s}_{2}\mathrm{s}}}\hphantom{h_{\mathrm{cc}_{2}}}=\frac{1}{N}\sum_{j=0}^{N-1}h_{j}\sin(2\phi_{j})\sin(\phi_{j})$
\State $\mathrlap{h_{\mathrm{s}_{2}\mathrm{c}}}\hphantom{h_{\mathrm{cc}_{2}}}=\frac{1}{N}\sum_{j=0}^{N-1}h_{j}\sin(2\phi_{j})\cos(\phi_{j})$
\State $\mathrlap{h_{\mathrm{sc}_{2}}}\hphantom{h_{\mathrm{cc}_{2}}}=\frac{1}{N}\sum_{j=0}^{N-1}h_{j}\sin(\phi_{j})\cos(2\phi_{j})$
\State $\mathrlap{h_{\mathrm{cc}_{2}}}\hphantom{h_{\mathrm{cc}_{2}}}=\frac{1}{N}\sum_{j=0}^{N-1}h_{j}\cos(\phi_{j})\cos(2\phi_{j})$
\columnbreak

\State $\mathrlap{\overline{v}}\hphantom{v_{\mathrm{cc}_{2}}}=\frac{1}{N}\sum_{j=0}^{N-1}v_{j}$
\State $\mathrlap{v_{\mathrm{s}}}\hphantom{v_{\mathrm{cc}_{2}}}=\frac{1}{N}\sum_{j=0}^{N-1}v_{j}\sin(\phi_{j})$
\State $\mathrlap{v_{\mathrm{c}}}\hphantom{v_{\mathrm{cc}_{2}}}=\frac{1}{N}\sum_{j=0}^{N-1}v_{j}\cos(\phi_{j})$ 
\State $\mathrlap{v_{\mathrm{ss}}}\hphantom{v_{\mathrm{cc}_{2}}}=\frac{1}{N}\sum_{j=0}^{N-1}v_{j}\sin^{2}(\phi_{j})$
\State $\mathrlap{v_{\mathrm{sc}}}\hphantom{v_{\mathrm{cc}_{2}}}=\frac{1}{N}\sum_{j=0}^{N-1}v_{j}\sin(\phi_{j})\cos(\phi_{j})$
\State $\mathrlap{v_{\mathrm{cc}}}\hphantom{v_{\mathrm{cc}_{2}}}=\frac{1}{N}\sum_{j=0}^{N-1}v_{j}\cos^{2}(\phi_{j})$

\State $\mathrlap{v_{\mathrm{s_{2}}}}\hphantom{v_{\mathrm{cc}_{2}}}=\frac{1}{N}\sum_{j=0}^{N-1}v_{j}\sin(2\phi_{j})$
\State $\mathrlap{v_{\mathrm{c_{2}}}}\hphantom{v_{\mathrm{cc}_{2}}}=\frac{1}{N}\sum_{j=0}^{N-1}v_{j}\cos(2\phi_{j})$
\State $\mathrlap{v_{\mathrm{s}_{2}\mathrm{s}}}\hphantom{v_{\mathrm{cc}_{2}}}=\frac{1}{N}\sum_{j=0}^{N-1}v_{j}\sin(2\phi_{j})\sin(\phi_{j})$
\State $\mathrlap{v_{\mathrm{s}_{2}\mathrm{c}}}\hphantom{v_{\mathrm{cc}_{2}}}=\frac{1}{N}\sum_{j=0}^{N-1}v_{j}\sin(2\phi_{j})\cos(\phi_{j})$
\State $\mathrlap{v_{\mathrm{sc}_{2}}}\hphantom{v_{\mathrm{cc}_{2}}}=\frac{1}{N}\sum_{j=0}^{N-1}v_{j}\sin(\phi_{j})\cos(2\phi_{j})$
\State $\mathrlap{v_{\mathrm{cc}_{2}}}\hphantom{v_{\mathrm{cc}_{2}}}=\frac{1}{N}\sum_{j=0}^{N-1}v_{j}\cos(\phi_{j})\cos(2\phi_{j})$
\end{algorithmic}
\end{multicols}

\begin{algorithmic}
\State $\mathrlap{s_\mathrm{w}}\hphantom{h_{\mathrm{cc}_{2}}}=\frac{(h_{\mathrm{s}_{2}\mathrm{c}}h_{\mathrm{sc}_{2}}-h_{\mathrm{s}_{2}\mathrm{s}}h_{\mathrm{cc}_{2}})(h_{\mathrm{cc}_{2}}h_{\mathrm{s_{2}}}-h_{\mathrm{s}_{2}\mathrm{c}}h_{\mathrm{c_{2}}})+(v_{\mathrm{s}_{2}\mathrm{c}}v_{\mathrm{sc}_{2}}-v_{\mathrm{s}_{2}\mathrm{s}}v_{\mathrm{cc}_{2}})(v_{\mathrm{cc}_{2}}v_{\mathrm{s_{2}}}-v_{\mathrm{s}_{2}\mathrm{c}}v_{\mathrm{c_{2}}})}{(h_{\mathrm{s}_{2}\mathrm{c}}h_{\mathrm{sc}_{2}}-h_{\mathrm{s}_{2}\mathrm{s}}h_{\mathrm{cc}_{2}})^{2}+(v_{\mathrm{s}_{2}\mathrm{c}}v_{\mathrm{sc}_{2}}-v_{\mathrm{s}_{2}\mathrm{s}}v_{\mathrm{cc}_{2}})^{2}}$
\State $\mathrlap{c_\mathrm{w}}\hphantom{h_{\mathrm{cc}_{2}}}=\frac{(h_{\mathrm{s}_{2}\mathrm{c}}h_{\mathrm{sc}_{2}}-h_{\mathrm{s}_{2}\mathrm{s}}h_{\mathrm{cc}_{2}})(h_{\mathrm{s}_{2}\mathrm{s}}h_{\mathrm{c_{2}}}-h_{\mathrm{sc}_{2}}h_{\mathrm{s_{2}}})+(v_{\mathrm{s}_{2}\mathrm{c}}v_{\mathrm{sc}_{2}}-v_{\mathrm{s}_{2}\mathrm{s}}v_{\mathrm{cc}_{2}})(v_{\mathrm{s}_{2}\mathrm{s}}v_{\mathrm{c_{2}}}-v_{\mathrm{sc}_{2}}v_{\mathrm{s_{2}}})}{(h_{\mathrm{s}_{2}\mathrm{c}}h_{\mathrm{sc}_{2}}-h_{\mathrm{s}_{2}\mathrm{s}}h_{\mathrm{cc}_{2}})^{2}+(v_{\mathrm{s}_{2}\mathrm{c}}v_{\mathrm{sc}_{2}}-v_{\mathrm{s}_{2}\mathrm{s}}v_{\mathrm{cc}_{2}})^{2}}$
\end{algorithmic}

\begin{multicols}{2}
\begin{algorithmic}
\State $\mathrlap{s_{\mathrm{h}}}\hphantom{h_{\mathrm{cc}_{2}}}=2(\mathrlap{s_{\mathrm{w}}h_{\mathrm{ss}}}\hphantom{s_{\mathrm{w}}h_{\mathrm{sc}}} +\mathrlap{c_{\mathrm{w}}h_{\mathrm{sc}}}\hphantom{c_{\mathrm{w}}h_{\mathrm{cc}}}+h_{\mathrm{s}})$
\State $\mathrlap{c_{\mathrm{h}}}\hphantom{h_{\mathrm{cc}_{2}}}=2(s_{\mathrm{w}}h_{\mathrm{sc}}+c_{\mathrm{w}}h_{\mathrm{cc}}+u_{\mathrm{c}})$
\State 
\State $\mathrlap{o_{\mathrm{h}}}\hphantom{h_{\mathrm{cc}_{2}}}=\hphantom{2(}\mathrlap{s_{\mathrm{w}}h_{\mathrm{s}}}\hphantom{s_{\mathrm{w}}h_{\mathrm{sc}}} +\mathrlap{c_{\mathrm{w}}h_{\mathrm{c}}}\hphantom{c_{\mathrm{w}}h_{\mathrm{cc}}}+\overline{h}$
\State $\mathrlap{a_{\mathrm{h}}}\hphantom{h_{\mathrm{cc}_{2}}}=\sqrt{s_{\mathrm{h}}^{2}+c_{\mathrm{h}}^{2}}$
\State $\mathrlap{\phi_{\mathrm{h}0}}\hphantom{h_{\mathrm{cc}_{2}}}=\arctan\!2(-c_{\mathrm{h}},s_{\mathrm{h}})$
\State
\State
\State
\columnbreak

\State $\mathrlap{s_{\mathrm{v}}}\hphantom{v_{\mathrm{cc}_{2}}}=2(\mathrlap{s_{\mathrm{w}}v_{\mathrm{ss}}}\hphantom{s_{\mathrm{w}}v_{\mathrm{sc}}} +\mathrlap{c_{\mathrm{w}}v_{\mathrm{sc}}}\hphantom{c_{\mathrm{w}}v_{\mathrm{cc}}}+v_{\mathrm{s}})$
\State $\mathrlap{c_{\mathrm{v}}}\hphantom{v_{\mathrm{cc}_{2}}}=2(s_{\mathrm{w}}v_{\mathrm{sc}}+c_{\mathrm{w}}v_{\mathrm{cc}}+v_{\mathrm{c}})$
\State 
\State $\mathrlap{o_{\mathrm{v}}}\hphantom{v_{\mathrm{cc}_{2}}}=\hphantom{2(}\mathrlap{s_{\mathrm{w}}v_{\mathrm{s}}}\hphantom{s_{\mathrm{w}}v_{\mathrm{sc}}} +\mathrlap{c_{\mathrm{w}}v_{\mathrm{c}}}\hphantom{c_{\mathrm{w}}v_{\mathrm{cc}}}+\overline{v}$
\State $\mathrlap{a_{\mathrm{v}}}\hphantom{v_{\mathrm{cc}_{2}}}=\sqrt{s_{\mathrm{v}}^{2}+c_{\mathrm{v}}^{2}}$
\State $\mathrlap{\phi_{\mathrm{v}0}}\hphantom{v_{\mathrm{cc}_{2}}}=\arctan\!2(-c_{\mathrm{v}},s_{\mathrm{v}})$
\State $\mathrlap{o_{\mathrm{w}}}\hphantom{v_{\mathrm{cc}_{2}}}=1$
\State $\mathrlap{a_{\mathrm{w}}}\hphantom{v_{\mathrm{cc}_{2}}}=\sqrt{s_{\mathrm{w}}^{2}+c_{\mathrm{w}}^{2}}$
\State $\mathrlap{\phi_{\mathrm{w}0}}\hphantom{v_{\mathrm{cc}_{2}}}=\arctan\!2(-c_{\mathrm{w}},s_{\mathrm{w}})$
\end{algorithmic}
\end{multicols}
\end{spacing}

\caption[Extraction of circular trajectories' sinusoid parameters as required
for Algorithm~\ref{alg:iterative-pmat-recon} from perspective projections.]{\label{alg:pmat-recon-sinusoids}Extraction of circular trajectories'
sinusoid parameters $a_{i\mathrm{h}}$, $a_{i\mathrm{v}}$, $a_{i\mathrm{w}}$,
$\phi_{i0\mathrm{h}}$, $\phi_{i0\mathrm{v}}$, $\phi_{i0\mathrm{w}}$,
$o_{i\mathrm{h}}$ and $o_{i\mathrm{v}}$ (cf.\ Eqs.~\ref{eq:sinusoid-fw-model_h}–\ref{eq:sinusoid-fw-model_v})
as required for Algorithm~\ref{alg:iterative-pmat-recon} from perspective
projections $(h_{i}(\phi),v_{i}(\phi))$, assuming an equidistant
sampling of the rotation phase $\phi\in[0,2\pi[$. The trajectory
index $i$ has been omitted for better readability. A detailed derivation
is given in Section~\ref{subsec:parameter-extraction}.}
\end{algorithm}
Now that the calibration problem has been formalized to the solution
of a small system of linear equations, an iterative scheme technique
shall be proposed for its self consistent solution. The calibration
equations (Eqs.\ \ref{eq:calib-a_h}–\ref{eq:calib-phi_w}) may to
this end be rearranged for each parameter to be determined, assuming
the respective parameters on the right-hand side known (beginning
with initial estimates). As there will be multiple analog equations
for each parameter relating it to either different rows of the projection
matrix or to multiple of the observed projected trajectories, weighted
averages will be used to adequately incorporate all available information
for each parameter. In order to avoid potential oscillations due to
inconsistencies, the iterative updates may be damped by a relaxation
factor $\lambda\in[0;1]$ scaling between no update ($\lambda=0$)
and no damping ($\lambda=1$). 

Without specifying the averaging weights yet, the described scheme
results in:
\begin{align*}
w_{i}:\! & =1\\
r_{i}'^{(k+1)} & =(1-\lambda)r_{i}'^{(k)}+\lambda\left\langle \frac{a_{im}}{P_{m\mathrm{a}}^{(k)}}\right\rangle _{m}\\
z_{i}'^{(k+1)} & =(1-\lambda)z_{i}'^{(k)}+\lambda\left\langle \frac{o_{im}-P_{m4}^{(k)}}{P_{m3}^{(k)}}\right\rangle _{m}\\
P_{m\mathrm{a}}^{(k+1)} & =(1-\lambda)P_{m\mathrm{a}}^{(k)}+\lambda\left\langle \frac{a_{im}}{r_{i}'^{(k)}}\right\rangle _{i}\\
\begin{array}{c}
P_{m3}^{(k+1)}\\
P_{m4}^{(k+1)}
\end{array}\negmedspace\negmedspace & =(1-\lambda)\negmedspace\!\begin{array}{c}
P_{m3}^{(k)}\\
P_{m4}^{(k)}
\end{array}\negmedspace\!+\lambda\:\underset{P_{m3}^{*},P_{m4}^{*}}{\mathrm{argmin}}\bigl(P_{m3}^{*}z_{i}'^{(k)}+P_{m4}^{*}-o_{im}\bigr)^{2}\\
\phi_{\mathrm{h}}:\! & =0\\
\phi_{\mathrm{v}} & =\left\langle \phi_{i0\mathrm{v}}-\phi_{i0\mathrm{h}}\right\rangle _{i}\\
\phi_{\mathrm{w}} & =\left\langle \phi_{i0\mathrm{w}}-\phi_{i0\mathrm{h}}\right\rangle _{i}
\end{align*}
where $k$ is the iteration index, $\bigl\langle\,\cdot\,\bigr\rangle_{i}$
and $\bigl\langle\,\cdot\,\bigr\rangle_{m}$ denote (to be further
defined) weighted averages over the marker index $i$ and the projection
matrix row index $m$ respectively. For uniformity of the representation,
the row related ``$\mathrm{h}$'', ``$\mathrm{v}$'' and ``$\mathrm{w}$''
indices of the observable sinusoid parameters are equally enumerated
by $m$ in the above averages, i.e.\ $\mathrm{h}\,\widehat{=}\,1,\mathrm{v}\,\widehat{=}\,2,\mathrm{w}\,\widehat{=}\,3$.
Based on the discussion given in Section~\ref{subsec:detector-tilt},
$w_{i}$ has been chosen $w_{i}=1$ without loss of generality. The
parameters $P_{m3},P_{m4}$ are determined by means of linear regression.
The phases $\phi_{\mathrm{w}}$ and $\phi_{\mathrm{v}}$ can be evaluated
independently as a weighted average over the available observations,
whereas $\phi_{\mathrm{h}}$ is, without loss of generality, defined
$0$.

Now that the iterative update scheme has been established, actual
averaging weights need to be defined. These weights shall respect
the varying certainty or error bar associated with the different available
equations for each parameter. It will be argued that the denominators
occurring in the right-hand-side expressions are reasonable indicators
in this respect and may be used as weighting factors, which in addition
avoids potential divisions by zero. This can be easily seen on the
example of $r_{i}'$: the observable amplitudes $a_{im}$ on the right
hand side correspond to actual horizontal and vertical ranges covered
by the projections observed on a rasterized detector. Given some absolute
precision in the determination of the projection's coordinates, the
relative uncertainty of some $a_{im}$ will be lower the larger its
absolute value. As $a_{im}$ are, given the relation $a_{im}=r_{i}'P_{m\mathrm{a}}$,
proportional to the parameters $P_{m\mathrm{a}}$, the latter may
as well be used for the respective importance weighting. This reasoning
can analogously be transferred to the remaining parameters as well,
 finally leading to the explicit alternating update scheme stated
in Algorithm~\ref{alg:iterative-pmat-recon}, whereby the weights
are formed as the square of the denominators to ensure positivity.
The averages over the differences $\phi_{i0\mathrm{v}}-\phi_{i0\mathrm{h}}$
and $\phi_{i0\mathrm{w}}-\phi_{i0\mathrm{h}}$ are performed (cf.\
Alg.~\ref{alg:iterative-pmat-recon}) over their complex amplitude
representation (with $\boldsymbol{i}$ denoting the imaginary unit).
The weighting factors $a_{i\mathrm{h}}^{2}$, $a_{i\mathrm{v}}^{2}$
and $a_{i\mathrm{w}}^{2}$ for the phases $\phi_{i0\mathrm{h}}$,
$\phi_{i0\mathrm{v}}$ and $\phi_{i0\mathrm{h}}$ thereby follow the
same heuristic as outlined previously.

Reasonable initial values for the trajectory parameters $r_{i}'$,
$z_{i}'$ and $w_{i}$ are
\begin{align}
r_{i}'^{(0)} & =a_{i\mathrm{h}}\label{eq:r_i_init}\\
z_{i}'^{(0)} & =o_{i\mathrm{v}}\label{eq:z_i_init}
\end{align}
which would also be the final results in case of a perfectly aligned
system. The projection matrix parameters $P_{m\mathrm{a}}$ may then
simply be initialized according to the intended iterative update scheme:
\begin{align}
P_{m\mathrm{a}}^{(0)} & =\frac{\sum_{i}r_{i}'^{(0)}a_{im}}{\sum_{i}r_{i}'^{(0)\,2}}\label{P_ma_init}\\
\shortintertext{\text{and \ensuremath{P_{m3}} and \ensuremath{P_{m4}} by a linear least squares fit}}P_{m3}^{(0)},P_{m4}^{(0)} & \underset{P_{m3},P_{m4}}{=\mathrm{argmin}}\left(P_{m3}z_{i}'^{(0)}+P_{m4}-o_{im}\right)
\end{align}

\subsection{Extraction of the trajectories' sinusoid parameters\label{subsec:parameter-extraction}}

As explained previously in Section~\ref{subsec:pmat-model}, arbitrary
cone beam projections of circular trajectories can be represented
in the following way:
\begin{align*}
u(\phi) & =\frac{a_{\mathrm{u}}\sin(\phi-\phi_{\mathrm{u}0})+o_{\mathrm{u}}}{a_{\mathrm{w}}\sin(\phi-\phi_{\mathrm{w}0})+1}\quad,
\end{align*}
with $u(\phi)$ on the left-hand side representing either of the projections'
$h(\phi)$ and $v(\phi)$ coordinates on the detection plane and $a_{\mathrm{u}},a_{\mathrm{w}},\phi_{\mathrm{u}0},\phi_{\mathrm{w}0},o_{\mathrm{u}}$
being the parameters encoding both the location of the projected marker
and properties of the projection geometry. While Section~\ref{subsec:pmat-model}
covered the reconstruction of both the actual trajectories and the
projection matrix from these parameters, their extraction from the
measured projections $(h(\phi),v(\phi))$ shall be detailed here using
an approach inspired by the Fourier decomposition of $h$ and $v$
used by Smekal et al 2004 \cite{Smekal2004}.

The above equation may be rearranged by multiplying with the denominator:
\begin{align*}
u(\phi)\left(a_{\mathrm{w}}\sin(\phi-\phi_{\mathrm{w}0})+1\right) & =a_{\mathrm{u}}\sin(\phi-\phi_{\mathrm{u}0})+o_{\mathrm{u}}
\end{align*}
and expanding the sine functions $a_{\mathrm{u}}\sin(\phi-\phi_{\mathrm{u}0})$
into $s_{\mathrm{u}}\sin(\phi)+c_{\mathrm{u}}\cos(\phi)$ such that
\begin{align}
u(\phi)\left(s_{\mathrm{w}}\sin(\phi)+c_{\mathrm{w}}\cos(\phi)+1\right) & =s_{\mathrm{u}}\sin(\phi)+c_{\mathrm{u}}\cos(\phi)+o_{\mathrm{u}}\label{eq:u-projection-relation-for-integration}
\end{align}
with
\begin{align*}
a_{\mathrm{u}} & =\sqrt{s_{\mathrm{u}}^{2}+c_{\mathrm{u}}^{2}}\\
\phi_{\mathrm{u}0} & =\arctan\!2\left(\sin(\phi_{\mathrm{u}0}),\cos(\phi_{\mathrm{u}0})\right)\\
 & =\arctan\!2(-c_{\mathrm{u}},s_{\mathrm{u}})\:.
\end{align*}

Now the above $\phi$ dependent equation (\ref{eq:u-projection-relation-for-integration})
relating $u(\phi)$ to the five parameters $s_{\mathrm{w}},c_{\mathrm{w}}$
and $s_{\mathrm{u}},c_{\mathrm{u}},o_{\mathrm{u}}$ can be expanded
into five $\phi$ independent equations by considering different integrals
of both sides exploiting the orthogonality relations of the sine and
cosine functions on the interval $\phi\in[0,2\pi)$. When considering
the right-hand side it is easy to see that integrating with respect
to $\phi$ over multiples of a period will effectively single out
$o_{\mathrm{u}}$, and similarly multiplying the equation by $\sin(\phi)$
or $\cos(\phi)$ prior to integration will ``select'' parameters
$s_{\mathrm{u}}$ or $c_{\mathrm{u}}$ respectively, while at the
same time obviously eliminating the $\phi$ dependence of the equation.
Further equations required to determine also $s_{\mathrm{w}}$ and
$c_{\mathrm{w}}$ are obtained by considering integrals over $\sin(2\phi)$
and $\cos(2\phi)$ respectively, which is equivalent to regarding
higher frequency components of the equations. Applying the aforesaid
integrals yields:
\[
\begin{array}{lllll}
s_{\mathrm{w}}\int_{0}^{2\pi}\negmedspace\mathrm{d}\phi\,u(\phi)\sin(\phi) & +\,c_{\mathrm{w}}\int_{0}^{2\pi}\negmedspace\mathrm{d}\phi\,u(\phi)\cos(\phi) & +\int_{0}^{2\pi}\negmedspace\mathrm{d}\phi\,u(\phi) & = & 2\pi o_{\mathrm{u}}\\
s_{\mathrm{w}}\int_{0}^{2\pi}\negmedspace\mathrm{d}\phi\,u(\phi)\sin^{2}(\phi) & +\,c_{\mathrm{w}}\int_{0}^{2\pi}\negmedspace\mathrm{d}\phi\,u(\phi)\sin(\phi)\cos(\phi) & +\int_{0}^{2\pi}\negmedspace\mathrm{d}\phi\,u(\phi)\sin(\phi) & = & \phantom{2}\pi s_{\mathrm{u}}\\
s_{\mathrm{w}}\int_{0}^{2\pi}\negmedspace\mathrm{d}\phi\,u(\phi)\sin(\phi)\cos(\phi) & +\,c_{\mathrm{w}}\int_{0}^{2\pi}\negmedspace\mathrm{d}\phi\,u(\phi)\cos^{2}(\phi) & +\int_{0}^{2\pi}\negmedspace\mathrm{d}\phi\,u(\phi)\cos(\phi) & = & \phantom{2}\pi c_{\mathrm{u}}\\
s_{\mathrm{w}}\int_{0}^{2\pi}\negmedspace\mathrm{d}\phi\,u(\phi)\sin(2\phi)\sin(\phi) & +\,c_{\mathrm{w}}\int_{0}^{2\pi}\negmedspace\mathrm{d}\phi\,u(\phi)\sin(2\phi)\cos(\phi) & +\int_{0}^{2\pi}\negmedspace\mathrm{d}\phi\,u(\phi)\sin(2\phi) & = & \phantom{2\pi}0\\
s_{\mathrm{w}}\int_{0}^{2\pi}\negmedspace\mathrm{d}\phi\,u(\phi)\cos(2\phi)\sin(\phi) & +\,c_{\mathrm{w}}\int_{0}^{2\pi}\negmedspace\mathrm{d}\phi\,u(\phi)\cos(2\phi)\cos(\phi) & +\int_{0}^{2\pi}\negmedspace\mathrm{d}\phi\,u(\phi)\cos(2\phi) & = & \phantom{2\pi}0\quad.
\end{array}
\]
When dividing by $2\pi$ and denoting the different integrals over
$u(\phi)$ by $\overline{u},u_{\mathrm{s}},u_{\mathrm{c}},u_{\mathrm{sc}}u_{\mathrm{ss}},u_{\mathrm{s}_{2}\mathrm{c}},\text{etc.}$
with indices indicating the accompanying sine and cosine terms (i.e.\
$u_{\mathrm{s}_{2}\mathrm{c}}=\frac{1}{2\pi}\int_{0}^{2\pi}\negmedspace\mathrm{d}\phi\,u(\phi)\sin(2\phi)\cos(\phi)$,
$\overline{u}=\frac{1}{2\pi}\int_{0}^{2\pi}\negmedspace\mathrm{d}\phi\,u(\phi)$),
the system of equations simplifies to
\begin{align}
s_{\mathrm{w}}u_{\mathrm{s}}+c_{\mathrm{w}}u_{\mathrm{c}}+\overline{u} & =o_{\mathrm{u}}\nonumber \\
s_{\mathrm{w}}u_{\mathrm{ss}}+c_{\mathrm{w}}u_{\mathrm{sc}}+u_{\mathrm{s}} & =\frac{s_{\mathrm{u}}}{2}\nonumber \\
s_{\mathrm{w}}u_{\mathrm{sc}}+c_{\mathrm{w}}u_{\mathrm{cc}}+u_{\mathrm{c}} & =\frac{c_{\mathrm{u}}}{2}\label{eq:param-eq-sys}\\
s_{\mathrm{w}}u_{\mathrm{s}_{2}\mathrm{s}}+c_{\mathrm{w}}u_{\mathrm{s}_{2}\mathrm{c}}+u_{\mathrm{s_{2}}} & =0\nonumber \\
s_{\mathrm{w}}u_{\mathrm{sc}_{2}}+c_{\mathrm{w}}u_{\mathrm{cc}_{2}}+u_{\mathrm{c_{2}}} & =0\quad,\nonumber 
\end{align}
with $s_{\mathrm{u}},c_{\mathrm{u}},o_{\mathrm{u}}$ and $s_{\mathrm{w}},c_{\mathrm{w}}$
being the sought unknowns. The latter two can be obtained by inverting
the last two equations, while the former can then be directly computed
using the first three equations:
\begin{align*}
\left[\begin{array}{cc}
u_{\mathrm{s}_{2}\mathrm{s}} & u_{\mathrm{s}_{2}\mathrm{c}}\\
u_{\mathrm{sc}_{2}} & u_{\mathrm{cc}_{2}}
\end{array}\right]\left[\begin{array}{c}
s_{\mathrm{w}}\\
c_{\mathrm{w}}
\end{array}\right] & =-\left[\begin{array}{c}
u_{\mathrm{s_{2}}}\\
u_{\mathrm{c_{2}}}
\end{array}\right]\\
\left[\begin{array}{c}
s_{\mathrm{w}}\\
c_{\mathrm{w}}
\end{array}\right] & =-\left[\begin{array}{cc}
u_{\mathrm{s}_{2}\mathrm{s}} & u_{\mathrm{s}_{2}\mathrm{c}}\\
u_{\mathrm{sc}_{2}} & u_{\mathrm{cc}_{2}}
\end{array}\right]^{-1}\left[\begin{array}{c}
u_{\mathrm{s_{2}}}\\
u_{\mathrm{c_{2}}}
\end{array}\right]\\
\left[\begin{array}{c}
s_{\mathrm{w}}\\
c_{\mathrm{w}}
\end{array}\right] & =(u_{\mathrm{s}_{2}\mathrm{c}}u_{\mathrm{sc}_{2}}-u_{\mathrm{s}_{2}\mathrm{s}}u_{\mathrm{cc}_{2}})^{-1}\left[\begin{array}{cc}
\hphantom{-}u_{\mathrm{cc}_{2}} & -u_{\mathrm{s}_{2}\mathrm{c}}\\
-u_{\mathrm{sc}_{2}} & \hphantom{-}u_{\mathrm{s}_{2}\mathrm{s}}
\end{array}\right]\left[\begin{array}{c}
u_{\mathrm{s_{2}}}\\
u_{\mathrm{c_{2}}}
\end{array}\right]
\end{align*}
i.e.
\begin{align}
s_{\mathrm{w}(u)} & =\frac{u_{\mathrm{cc}_{2}}u_{\mathrm{s_{2}}}-u_{\mathrm{s}_{2}\mathrm{c}}u_{\mathrm{c_{2}}}}{u_{\mathrm{s}_{2}\mathrm{c}}u_{\mathrm{sc}_{2}}-u_{\mathrm{s}_{2}\mathrm{s}}u_{\mathrm{cc}_{2}}}\label{eq:sinusoid-params-start}\\
c_{\mathrm{w}(u)} & =\frac{u_{\mathrm{s}_{2}\mathrm{s}}u_{\mathrm{c_{2}}}-u_{\mathrm{sc}_{2}}u_{\mathrm{s_{2}}}}{u_{\mathrm{s}_{2}\mathrm{c}}u_{\mathrm{sc}_{2}}-u_{\mathrm{s}_{2}\mathrm{s}}u_{\mathrm{cc}_{2}}}\\
o_{\mathrm{u}} & =\hphantom{2(}s_{\mathrm{w}}u_{\mathrm{s}}+c_{\mathrm{w}}u_{\mathrm{c}}+\overline{u}\label{eq:sinusoid-params-o_u}\\
s_{\mathrm{u}} & =2(s_{\mathrm{w}}u_{\mathrm{ss}}+c_{\mathrm{w}}u_{\mathrm{sc}}+u_{\mathrm{s}})\\
c_{\mathrm{u}} & =2(s_{\mathrm{w}}u_{\mathrm{sc}}+c_{\mathrm{w}}u_{\mathrm{cc}}+u_{\mathrm{c}})\\
a_{\mathrm{u}} & =\sqrt{s_{\mathrm{u}}^{2}+c_{\mathrm{u}}^{2}}\\
a_{\mathrm{w}} & =\sqrt{s_{\mathrm{w}}^{2}+c_{\mathrm{w}}^{2}}\label{eq:sinusoid-params-a_w}\\
\phi_{\mathrm{w}0} & =\arctan\!2(-c_{\mathrm{w}},s_{\mathrm{w}})\\
\phi_{\mathrm{u}0} & =\arctan\!2(-c_{\mathrm{u}},s_{\mathrm{u}})\quad.
\end{align}

Although all five parameters can be determined independently for both
horizontal and vertical projection components $h(\phi)$ and $v(\phi)$
respectively (represented by $u(\phi)$ here), $s_{\mathrm{w}}$ and
$c_{\mathrm{w}}$ – or equivalently $a_{\mathrm{w}}$ and $\phi_{\mathrm{w}0}$
– are shared parameters that are expected to be identical for both
$h(\phi)$ and $v(\phi)$. It is therefore advisable to use a weighted
average of the respective $s_{\mathrm{w}(u)}$ and $c_{\mathrm{w}(u)}$
parameters obtained from the horizontal ($u=\mathrm{h}$) and vertical
($u=\mathrm{v}$) projection components for the computation of the
remaining parameters. A sensible choice for the relative importance
weights are the respective determinants of both available systems
of equations for $s_{w}$ and $c_{w}$, i.e.:
\begin{align}
s_{w} & =\frac{(h_{\mathrm{s}_{2}\mathrm{c}}h_{\mathrm{sc}_{2}}-h_{\mathrm{s}_{2}\mathrm{s}}h_{\mathrm{cc}_{2}})(h_{\mathrm{cc}_{2}}h_{\mathrm{s_{2}}}-h_{\mathrm{s}_{2}\mathrm{c}}h_{\mathrm{c_{2}}})+(v_{\mathrm{s}_{2}\mathrm{c}}v_{\mathrm{sc}_{2}}-v_{\mathrm{s}_{2}\mathrm{s}}v_{\mathrm{cc}_{2}})(v_{\mathrm{cc}_{2}}v_{\mathrm{s_{2}}}-v_{\mathrm{s}_{2}\mathrm{c}}v_{\mathrm{c_{2}}})}{(h_{\mathrm{s}_{2}\mathrm{c}}h_{\mathrm{sc}_{2}}-h_{\mathrm{s}_{2}\mathrm{s}}h_{\mathrm{cc}_{2}})^{2}+(v_{\mathrm{s}_{2}\mathrm{c}}v_{\mathrm{sc}_{2}}-v_{\mathrm{s}_{2}\mathrm{s}}v_{\mathrm{cc}_{2}})^{2}}\label{eq:sinusoid-params-mean-s_w}\\
c_{w} & =\frac{(h_{\mathrm{s}_{2}\mathrm{c}}h_{\mathrm{sc}_{2}}-h_{\mathrm{s}_{2}\mathrm{s}}h_{\mathrm{cc}_{2}})(h_{\mathrm{s}_{2}\mathrm{s}}h_{\mathrm{c_{2}}}-h_{\mathrm{sc}_{2}}h_{\mathrm{s_{2}}})+(v_{\mathrm{s}_{2}\mathrm{c}}v_{\mathrm{sc}_{2}}-v_{\mathrm{s}_{2}\mathrm{s}}v_{\mathrm{cc}_{2}})(v_{\mathrm{s}_{2}\mathrm{s}}v_{\mathrm{c_{2}}}-v_{\mathrm{sc}_{2}}v_{\mathrm{s_{2}}})}{(h_{\mathrm{s}_{2}\mathrm{c}}h_{\mathrm{sc}_{2}}-h_{\mathrm{s}_{2}\mathrm{s}}h_{\mathrm{cc}_{2}})^{2}+(v_{\mathrm{s}_{2}\mathrm{c}}v_{\mathrm{sc}_{2}}-v_{\mathrm{s}_{2}\mathrm{s}}v_{\mathrm{cc}_{2}})^{2}}\,.\label{eq:sinusoid-params-mean-c_w}
\end{align}
The weighting accounts for the generally considerably differing amplitudes
of $h(\phi)$ and $v(\phi)$ and the therefore differing relative
error within the derived quantities; and in particular also for the
singular case of one of the determinants becoming zero. This occurs
in the case of a projection view parallel to the circular trajectory
when either $h(\phi)$ or $v(\phi)$ become constant ($\phi$ independent),
which commonly is the case for $v(\phi)$ in a perfectly aligned system.

Algorithm~\ref{alg:pmat-recon-sinusoids} compiles the above derivations
into an explicit procedure for the deduction of the sinusoid parameters
$a_{i\mathrm{h}}$, $a_{i\mathrm{v}}$, $a_{i\mathrm{w}}$, $\phi_{i0\mathrm{h}}$,
$\phi_{i0\mathrm{v}}$, $\phi_{i0\mathrm{w}}$, $o_{i\mathrm{h}}$
and $o_{i\mathrm{v}}$ fully describing the perspective projections
of rotating points (cf.\ Eqs.~\ref{eq:sinusoid-fw-model_h}–\ref{eq:sinusoid-fw-model_v}).
The inner products with respect to various trigonometric functions
are expressed as explicit sums over equidistantly sampled rotation
angles $\phi_{j}$. The nature of trigonometric functions and the
periodic form of the considered problem thereby ensures that no discretization
errors are introduced, analog to classic Fourier analysis (cf.\ e.g.\
the Handbook of Mathematics \cite{Bronstein2001}). Indeed, these
sums can, by application of trigonometric identities, also be formulated
in terms of standard Fourier coefficients $u_{\mathrm{s}_{n}}=\frac{1}{N}\sum_{j=0}^{N-1}u_{j}\sin(n\frac{2\pi}{N}j)$
and $u_{\mathrm{c}_{n}}=\frac{1}{N}\sum_{j=0}^{N-1}u_{j}\cos(n\frac{2\pi}{N}j)$,
such that e.g.\ $u_{\mathrm{sc}_{2}}=\frac{1}{4}(u_{\mathrm{s}_{1}}+u_{\mathrm{s}_{3}})$,
with $u\in\{h,v\}$.

The latter representation reveals that Fourier components of $u(\phi)$
up to the third harmonic are implicitly used in the above derivations
implying that $N\ge6$ (despite the fact that only 5 parameters are
actually to be determined), which is e.g.\ also consistent with the
publication by Noo et al.\ \cite{Noo2000} in context of a completely
different method  for the actual retrieval of the projection parameters.
Fourier coefficients up to the third harmonic of the projected trajectories
were also used by Smekal et al.\ \cite{Smekal2004} as input to their
calibration procedure.

\bibliographystyle{plain}
\bibliography{}

\end{document}